\documentclass{article}
\usepackage{eurosym}
\usepackage{amsmath}
\usepackage{amssymb}
\usepackage{setspace}
\usepackage{color,ulem}
\usepackage{setspace}

\newtheorem{theorem}{Theorem}

\newtheorem{claim}{Claim}

\newtheorem{definition}{Definition}

\newtheorem{lemma}{Lemma}

\newtheorem{proposition}{Proposition}
\newtheorem{remark}{Remark}

\begin{document}

\title{A Stable and Strategy-Proof \\
Controlled School Choice Mechanism\\
with Integrated and Flexible Rules\thanks{%
This paper is a revised version of a part of our unpublished paper, entitled
\textquotedblleft Stable Mechanisms in Controlled School
Choice\textquotedblright , with additional modifications. This work was
supported by JSPS KAKENHI Grant Number 22K01402, JP20K01675, JP19K01542 and
JP18K01513.}}
\author{Minoru Kitahara\thanks{%
Department of Economics, Osaka Metropolitan University} \and Yasunori Okumura%
\thanks{%
Corresponding author. Department of Logistics and Information Engineering,
TUMSAT, 2-1-6, Etchujima, Koto-ku, Tokyo, 135-8533 Japan.
Phone:+81-3-5245-7300. Fax:+81-3-5245-7300. E-mail:
okuyasu@gs.econ.keio.ac.jp}}
\maketitle

\begin{center}
\textbf{Abstract}
\end{center}

We examine a controlled school choice model where students are categorized
into different types, and the distribution of these types within a school
influences its priority structure. This study provides a general framework
that integrates existing controlled school choice models, including those
utilizing reserve rules, quota rules, and bonus-point rules. Specifically,
we introduce an adjusted scoring rule that unifies these rules. By achieving
a matching that satisfies the stability defined in this framework, matching
authorities can effectively manage the trade-offs inherent in controlled
school choice markets. Moreover, the priority order for a school is
represented as a weak order with each given assignment, meaning that ties
are allowed. Our mechanism ensures a stable matching and satisfies
strategy-proofness. In particular, when priority orders are restricted to
linear orders with each given assignment, our mechanism guarantees
student-optimal stability.

\textbf{Keywords}: Controlled school choice; Affirmative action; Adjusted
scoring; Stable mechanism; Strategy-proofness

\textbf{JEL Classification Numbers}: C78; D47 \newpage

\section{Introduction}

Many real-world controlled school admission policies aim to foster diversity
and support socioeconomically disadvantaged students through affirmative
action policies. Examples include the reserves rule, used in Chicago public
high schools, which sets aside a certain number of seats for disadvantaged
students (Dur et al., 2020), and the priority-based rule in China, which
grants minority students a score bonus in entrance exams (Wang, 2009).
However, as explained later in this section, implementing such policies
often involves a trade-off between promoting diversity and managing
associated costs, making it difficult to operate them appropriately.

In this study, we introduce a general class of priority rules that allows
for a flexible and integrated approach to existing affirmative action
measures, such as reserves, quotas, and bonus-point rules. Specifically, our
priority rule takes into account the existing student distribution in a
school when determining the priority order of unassigned students. Moreover,
unlike traditional models that assume a strict ranking of students, the
priority order for school is represented as a weak order, which allows for
ties in priority rankings with a given type distribution.

For example, consider two subsets of students, $J$ and $J^{\prime }$, each
missing one student to reach the full capacity of a school. Let $i$ and $j$
be students who are not in either $J$ or $J^{\prime }$. Suppose that under
the school's priority rule, $i$ has a higher priority than $j$ for the
remaining seat when added to $J$. On the other hand, $J^{\prime }$ contains
more students of the same type as $i$ and/or fewer of the same type as $j$.
Then, $j$ may have a higher priority than $i$ for the last seat when added
to $J^{\prime }$.

In our model, a matching violates priority (or has justified envy) if there
exists a student $i$ who prefers a school $s$ over their assigned school,
and there also exists student $j$ who is assigned to $s$ and has a lower
priority than $i$ with \textit{the set of assigned students in }$s$\textit{\
excluding only }$j$. If no such student-school pair exists, the matching is
considered fair. This definition is equivalent to the usual one as long as
the priority order of students does not change regardless of the subset of
students given.

By adopting such a notion of fairness in matchings, we can flexibly address
the goal of promoting diversity and its associated costs. The mechanism
introduced in this study ensures a fair (more strictly stable) matching in
this sense. Moreover, our mechanism is (group) strategy-proof, meaning that
no student (group of students) has an incentive to misreport their
preferences.

Through these existing rules, a disadvantaged student will gain admission to
a high-quality school, while an advantaged student will not, despite both
expressing a desire to enter the school and the former having a lower
entrance examination score than the latter. However, high-quality schools
often require extensive academic preparation, and students with lower test
scores are often insufficiently prepared. Thus, admitting a (disadvantaged)
student with a low score may increase the costs for school staff and other
students in the school, because additional support may be necessary for the
disadvantaged student to keep up, and/or adjustments to the existing program
may be required. Moreover, some empirical studies, such as those reviewed by
Arcidiacono and Lovenheim (2016), note that such admission policies may harm
the disadvantaged themselves, if a program is more difficult than they
expected.\footnote{%
See, for example, Arcidiacono et al. (2011, 2016) and Sander and Taylor
(2012) on this issue. Arcidiacono and Lovenheim (2016) state that while
there is no clear evidence showing a negative effect on the likelihood of
graduation for minority students, some evidences suggest that
less-academically prepared minority students are likely to change their
major to a less demanding major. Moreover, if such students had gone to a
lower-level school, then they would have had the chance to graduate with a
more difficult major.}

Therefore, schools and/or matching authorities face a trade-off between
enhancing diversity or providing a better environment for disadvantaged
students and reducing the associated costs expected to be associated with
low test scores for both the school and the student.\footnote{%
Recently, Imamura (2023) also notes a similar trade-off regarding
affirmative action.} Then, we consider adopting the reserves rule introduced
earlier. To implement this rule, the number of seats reserved for
disadvantaged students must be determined before students submit their
applications. The number of seats in a school reserves may be excessive if
the entrance test scores of disadvantaged applicants are overall lower than
expected because the applicants' scores reflect their level of preparedness
for a school's curriculum of that school. Conversely, in the opposite
scenario, the number of reserved seats may be insufficient. Thus, selecting
the appropriate number of reserved seats for disadvantaged students in
advance may be hard.

Similar problems arise with other rules, such as the (advantaged) quota and
the priority-based rules. In the former, the quota level should be
determined before students submit applications, whereas in the latter rule,
the bonus-point levels for disadvantaged students should be determined
beforehand. However, under both rules, selecting the appropriate levels
requires information on the preferences and scores of applicants.

As a potential solution to this problem, we introduce an adjusted scoring
rule that integrates existing rules, along with a stable and strategy-proof
mechanism that operates under this rule. For example, the following rule
could be implemented: If a school has already admitted $n$ or more
disadvantaged students, the bonus-point level for a newly applying
disadvantaged student is set to zero; otherwise, the level is set at $x$ ($%
x>0$).

The mechanisms introduced in previous studies, which are discussed in
Section 2, cannot flexibly account for trade-offs, as they are designed to
handle only specific rules, such as the reserves rule. In contrast, we
establish a strategy-proof and stable mechanism that can adapt to trade-offs
more flexibly by incorporating general priority rules, including the
adjusted scoring rule. The weak order structure of our priority rules
further ensures that ties in priority rankings are handled systematically,
enhancing the practicality of our approach.

\section{Literature Review}

This study considers a general controlled school choice model. This field is
first discussed by Abdulkadiro\u{g}lu and S\"{o}nmez (2003) and Abdulkadiro%
\u{g}lu (2005). Since then, various rules, including reserve- and/or
quota-based rules and priority-based rules, are extensively studied in
numerous papers. Our rule serves as a hybrid, incorporating elements of
them. In the following, we explain how our rule integrates these existing
frameworks and findings.

Echenique and Yenmez (2015) note that aiming for a diverse student body in a
school may make the school's choice of students complementaries. However,
several previous studies such as Hafalir et al. (2013), Ehlers et al.
(2014), Echenique and Yenmez (2015), Erdil and Kumano (2019), and Imamura
(2023) have succeeded in showing that several applicative choice rules
reflecting such aims satisfy substitutability and thus a stable matching can
simply be attained by the SPDA algorithm. In our study, we also introduce a
method to construct a choice function that satisfies substitutability from
one in the class of general priority rules that includes the adjusted
scoring rule.

Ehlers et al. (2014) introduce both soft and hard bounds rules,
incorporating upper and lower limits for each type of students and each
school. The soft bounds rule is a generalization of that introduced by
Hafalir et al. (2013). The adjusted scoring rule includes the soft bounds
rule along with more complex variations.\footnote{%
Westkamp (2013) and Ayg\"{u}n and Turhan (2020) also consider kinds of the
soft bounds models. In their models, there is a target distribution and if
it is not achieved because too few students with a type that apply to a
school, then the capacity of the school and the type is redistributed in
accordance with a capacity transfer scheme. Our model is not a
generalization of their models, since then a priority between two different
type students may be affected by a redistribution of another type capacity,
which violates one of our properties.} For example, the occupation degrees
of the types of students in a school is represented by arbitrary $n$ levels,
and if the level of a type of current assignment in the school is higher
than that of another type, then a student of the latter type has a higher
priority in the school than one of the former type in the school. Moreover,
although our model is not a direct generalization of the hard bounds model,
we can consider hard upper bounds in the adjusted scoring rule.

Imamura (2023) introduces the reserves-and-quotas rule, which includes rules
introduced by Echenique and Yenmez (2015) and can be regarded as a rule with
hard upper bound\ and soft lower bound in the sense of Ehlers et al. (2014).
The adjusted scoring rule also includes the reserves-and-quotas rule.
Imamura (2023) also considers a trade-off in adopting such a rule, called
the trade-off between meritocracy and diversity. On the meritocracy side, he
focuses only on the priority ranking of students. However, with our rule, we
can consider not only the priority ranking of students but also the
quantitative score differences among them.

Erdil and Kumano (2019) introduce the rule with the target distributions of
types in schools, which is included in the adjusted scoring rule.\footnote{%
Erdil and Kumano (2019) also provide a model called admissions by a
committee, that is beyond the scope of our model.} Moreover, unlike previous
models, the priority orders for schools are allowed to be a weak order in
their model. Alternatively, it can be said that they generalize the model
presented by Erdil and Ergin (2008) with distributional constraints. In our
model, we also allow weak orders priorities. Therefore, by employing our
rule, we can expand on various rules mentioned above to accommodate
priorities for schools as weak orders.

Moreover, we successfully construct a choice function that satisfies size
monotonicity and substitutability from the priority rule that includes the
adjusted scoring rule. Thus, by using the functions, we introduce a stable
and strategy-proof mechanism. This implies that previous finding on simple
strategy-proof and stable mechanisms are extended with ties, which is an
answer (limited to our model) to the question of Erdil and Kumano (2019)
regarding the existence of a strategy-proof SPDA-like algorithm in more
general settings.

Our rules can also be seen as an extension of the priority-based rule, which
is firstly introduced by Kojima (2012) and widely examined in previous
studies. For a survey of these studies, see, for example, Kitahara and
Okumura (2023b). The adjusted scoring rule allows the priority ranking to be
modified based on the distribution of student types assigned. Thus, our rule
serves as a hybrid, incorporating elements of the reserve- and/or
quota-based rules and the priority-based rule.

S\"{o}nmez and Switzer (2013) and Kominers and S\"{o}nmez (2016) discuss
slot-specific priority models, where each school has multiple slots, each
with its own priority order. Our approach appears similar to the
slot-specific priority model; however, it is completely independent. In
fact, as shown in Appendix E, there exist cases where the matching achieved
by our approach cannot be replicated using slot-specific priorities.

Finally, Kitahara and Okumura (2023a) propose a method to derive a
student-optimal stable matching based on the algorithm introduced by Erdil
and Ergin (2008) when schools' priority orders are allowed to be weak. To
implement their method, an initial stable matching must first be obtained.
In our study, the mechanism guarantees a student-optimal stable matching
only when all schools have linear priority orders with any given
assignments. If any school's priority order is a weak order rather than a
linear order with some assignments, our mechanism alone may be insufficient
to achieve a student-optimal stable matching. Therefore, by combining the
method proposed by Kitahara and Okumura (2023a) with our mechanism, we can
ensure student-optimal stable matchings even when some schools do not have
linear priority orders.

\section{Basic Model}

Let $S$ and $I$ be finite sets of schools and students, respectively. Let $%
q_{s}\geq 1$ denote the capacity of school $s,$ which is arbitrarily fixed
for all $s\in S$. A \textbf{matching} is a mapping $\mu $ satisfying

\begin{description}
\item[(1)] $\mu \left( i\right) $ $\in S\cup \{\emptyset \}$ for all $i\in I$
and $\mu \left( s\right) \subseteq I$ for all $s\in S$,

\item[(2)] $\mu \left( i\right) =s\in S$ if and only if $i\in \mu \left(
s\right) $,

\item[(3)] $\mu \left( i\right) =\emptyset $ if and only if $i\notin \mu
\left( s\right) $ for all $s\in S$,

\item[(4)] $\left\vert \mu \left( s\right) \right\vert \leq q_{s}$ for all $%
s\in S$.
\end{description}

Note that $\mu \left( i\right) =\emptyset $ means that student $i$ is not
assigned to any school for $\mu $.

A finite type space is given by $T$. Let $\tau ^{s}:I\cup \left\{ \emptyset
\right\} \rightarrow T$ be a type function for school $s$, where $\tau
^{s}\left( i\right) $ is the type of student $i$ based on the criterion of $%
s $. For notational convenience, let $\tau ^{s}\left( \emptyset \right) \neq
\tau ^{s}\left( i\right) $ for all $i\in I$ and all $s\in S$. Models used in
the previous studies assume $\tau ^{s}=\tau ^{s^{\prime }}$ for all $%
s,s^{\prime }\in S$. However, we generalize that the criteria of schools
with regard to \textquotedblleft types\textquotedblright\ vary among them.
For example, even if students $i$ and $j$ are in different type groups
according to the criterion of $s;$ that is, $\tau ^{s}\left( i\right) \neq
\tau ^{s}\left( j\right) $, they can have the same type according to the
criterion of $s^{\prime }$; that is, $\tau ^{s^{\prime }}\left( i\right)
=\tau ^{s^{\prime }}\left( j\right) $.\footnote{%
For example, school $s$ divides the students into two income types (i.e.,
low-income and high-income students), but school $s^{\prime }$ divides the
students into three income types (i.e., low-income, middle-income and
high-income students). Then, students $i$ and $j$, who are the middle-income
type according to the criterion of $s^{\prime }$, can be a high-income
student and a low-income student according to the criterion of $s$,
respectively.}

Each student has only one type for $s$. For $t\in T$, let $I_{s}^{t}$ be a
set of type $t$ students for school $s$; that is, $i\in I_{s}^{t}$ implies $%
\tau ^{s}\left( i\right) =t$. We assume that for all $s\in S$, $%
I_{s}^{t}\cap I_{s}^{t^{\prime }}=\emptyset $ for all $t\neq t^{\prime }$
and $\cup _{t\in T}I_{s}^{t}=I$. Since there is no fear of misunderstanding,
hereafter, we omit $s$ from $\tau ^{s}$ and $I_{s}^{t}$.

Let $\mathcal{P}$ be the set of all linear orders (complete, transitive and
antisymmetric binary relation) over $S\cup \{\emptyset \}$. Let $P=\left(
P_{i}\right) _{i\in I}\in \mathcal{P}^{\left\vert I\right\vert }$ be the
profile of preference relations for all students, where $P_{i}\in \mathcal{P}
$ represents the preference relation of $i\in I$. Let $sP_{i}s^{\prime }$
mean that $i$ prefers $s$ to $s^{\prime }$. We denote $sR_{i}s^{\prime }$ if 
$sP_{i}s^{\prime }$ or $s=s^{\prime }$. A school $s$ is \textbf{acceptable}
for $i$ if $sR_{i}\emptyset $.

Let $\succsim _{s}^{J}$ be a weak order (complete and transitive binary
relation) with 
\begin{equation*}
J\in \mathcal{J}_{s}=\left\{ \left. J^{\prime }\subseteq I\text{ }%
\right\vert \text{ }\left\vert J^{\prime }\right\vert \leq q_{s}-1\right\}
\end{equation*}%
for school $s\in S$ over $\left( I\setminus J\right) \cup \left\{ \emptyset
\right\} $. This is the priority order for $s$ when the set of assigned
students in $s$ is \textit{given by} $J$ with $\left\vert J\right\vert \leq
q_{s}-1$; that is, at least one seat is not filled in $s$. As usual, we let $%
\succsim _{s}^{J}$ be the asymmetric part and $\sim _{s}^{J}$ be the
symmetric part of $\succsim _{s}^{J}$.

For $J\in \mathcal{J}_{s}$ and $i,j\in \left( I\setminus J\right) \cup
\left\{ \emptyset \right\} ,$ $i\succ _{s}^{J}j$ indicates that $i$ takes
priority over $j$ in the case where the set of students $J\in \mathcal{J}%
_{s} $ is assigned to $s$. In that case, we simply state that $i$ (resp. $j$%
) has a \textbf{lower (}resp. \textbf{higher) priority }than $j$ (resp. $i$)
with $J$. On the other hand, $i\sim _{s}^{J}j$ implies that they are tied in 
$s$ with $J\in \mathcal{J}_{s}$. If $i\succsim _{s}^{J}\emptyset $, then
student $i$ is said to be \textbf{acceptable} for $s$ with $J$.

Let $\succsim _{s}=\left( \succsim _{s}^{J}\right) _{J\in \mathcal{J}_{s}}$
be a \textbf{priority order profile} for $s$ and $\succsim =\left( \succsim
_{s}\right) _{s\in S}$.

As formally defined later, we allow a school's priority order to depend on
its current assignment. That is, $i\succsim _{s}^{J}j$ and $j\succ
_{s}^{J^{\prime }}i$ can both hold, depending on the type distributions in $%
J $ and $J^{\prime }$. Specifically, type $\tau ^{s}\left( i\right) $
students are relatively rare compared to type $\tau ^{s}\left( j\right) $
students in $J$, and vice versa in $J^{\prime }$.

In particular, suppose that 
\begin{equation}
\succsim _{s}^{J}\cap \left( J\cap J^{\prime }\right) \times \left( J\cap
J^{\prime }\right) =\succsim _{s}^{J^{\prime }}\cap \left( J\cap J^{\prime
}\right) \times \left( J\cap J^{\prime }\right)  \label{b}
\end{equation}%
for all $J,J^{\prime }\in \mathcal{J}_{s}$ and all $s\in S$. Then, our model
is equivalent to that of Erdil and Ergin (2008). Moreover, if $\succsim
_{s}^{J}$ is a linear orders for all $J\in \mathcal{J}_{s}$ and all $s\in S$
and (\ref{b}) is satisfied for all $J,J^{\prime }\in \mathcal{J}_{s}$ and
all $s\in S$, then our model is equivalent to the usual many-to-one matching
model introduced by Gale and Shapley (1962).

We define some properties on matchings. Fix $(P,\succsim )$. A matching $\mu 
$ is \textbf{non-wasteful} if for any student $i$ and any school $s$, $%
sP_{i}\mu \left( i\right) $ implies $\emptyset \succ _{s}^{\mu \left(
s\right) }i$ or $\left\vert \mu \left( s\right) \right\vert =q_{s}$. A
matching $\mu $ is \textbf{individually rational} if $\mu \left( i\right)
R_{i}\emptyset $ for all $i\in I$ and $i\succsim _{s}^{\mu \left( s\right)
\setminus \{i\}}\emptyset $ for all $s$ and all $i\in \mu \left( s\right) $.

A student $i$ has \textbf{justified envy} toward $j$ at $\mu $ if $s=\mu
\left( j\right) P_{i}\mu \left( i\right) $ and $i\succ _{s}^{\mu \left(
s\right) \setminus \left\{ j\right\} }j$. A matching $\mu $ is \textbf{fair}%
\textit{\ }if no student has justified envy at $\mu $. A matching $\mu $ is 
\textbf{stable} if it is non-wasteful, individually rational and fair. A
matching $\mu $ \textbf{Pareto dominates} $\mu ^{\prime }$ if $\mu \left(
i\right) R_{i}\mu ^{\prime }\left( i\right) $ for all $i\in I$ and $\mu
\left( i\right) P_{i}\mu ^{\prime }\left( i\right) $ for some $i\in I$.
Moreover, a matching $\mu $ is \textbf{student-optimal stable} if $\mu $ is
stable and is not Pareto dominated by any other stable matchings.

Note that if (\ref{b}) is satisfied for all $J,J^{\prime }\in \mathcal{J}%
_{s} $ and all $s\in S$, these properties are the same as those of many
previous studies.

Next, we introduce properties on priority orders. Let $i,j,k\in I\cup
\left\{ \emptyset \right\} $, $s\in S$ and $J,J^{\prime }\subseteq I$.

\begin{description}
\item[PD (Preference for Diversity)] Suppose $i,j\notin J$, $i,j\notin
J^{\prime }$, $\left\vert J\cap I^{\tau \left( i\right) }\right\vert \geq
\left\vert J^{\prime }\cap I^{\tau \left( i\right) }\right\vert $ and $%
\left\vert J\cap I^{\tau \left( j\right) }\right\vert \leq \left\vert
J^{\prime }\cap I^{\tau \left( j\right) }\right\vert $. Then, $i\succ
_{s}^{J}j$ implies $i\succ _{s}^{J^{\prime }}j$ and $i\succsim _{s}^{J}j$
implies $i\succsim _{s}^{J^{\prime }}j$.

\item[WO (Within-Type Ordering)] Suppose $i,j\notin J$, $i,j\notin J^{\prime
}$, and $\tau \left( i\right) =\tau \left( j\right) $. Then, $i\succ
_{s}^{J}j$ implies $i\succ _{s}^{J^{\prime }}j$ and $i\succsim _{s}^{J}j$
implies $i\succsim _{s}^{J^{\prime }}j$.

\item[DC (Distributional Consistency of WO)] Suppose $i\neq j,$ $i,j,k\notin
J$, $\tau \left( i\right) =\tau \left( j\right) $ and $i\succsim _{s}^{J}j$.
Then, $j\succ _{s}^{J\cup \left\{ i\right\} }k$ implies $i\succ _{s}^{J\cup
\left\{ j\right\} }k,$ $j\sim _{s}^{J\cup \left\{ i\right\} }k$ implies $%
i\sim _{s}^{J\cup \left\{ j\right\} }k$, and $k\succ _{s}^{J\cup \left\{
i\right\} }j$ implies $k\succ _{s}^{J\cup \left\{ j\right\} }i$.
\end{description}

First, we explain \textbf{PD}.\ Suppose $i\succ _{s}^{J}j$; that is, $i$
takes priority over $j$ for filling the seat for $s$ to add $J$. We consider
another assignment $J^{\prime }\in \mathcal{J}_{s}$ such that the number of
type $\tau \left( i\right) $ students does not increase and that of $\tau
\left( j\right) $-type students does not decrease compared with those in $J$%
. Then, $i\succ _{s}^{J^{\prime }}j$; that is, $i$ still takes priority over 
$j$ for filling the seat for $s$.

Note that in \textbf{PD}, either $i$ or $j$ can be $\emptyset $. That is, if 
$i\notin J\cup J^{\prime }$ and $\left\vert J\cap I^{\tau \left( i\right)
}\right\vert \geq \left\vert J^{\prime }\cap I^{\tau \left( i\right)
}\right\vert $, then $i\succsim _{s}^{J}\emptyset $ implies $i\succsim
_{s}^{J^{\prime }}\emptyset $ (resp. $\emptyset \succ _{s}^{J^{\prime }}i$
implies $\emptyset \succ _{s}^{J}i$). This implies that any decrease (resp.
increase) in the type $\tau \left( i\right) $ does not make $i$ unacceptable
(acceptable).

In the school choice models in most previous studies, the acceptability of
each student is independent of who other students are. This rules out the
case in which a student becomes acceptable if her type is so rare that her
diversity value prevails over her low expected academic preparation.
However, we allow for the acceptability of a school for a student also
potentially being dependent on the type distribution of the students
assigned to the school to flexibly manage the trade-off introduced in
Section 1. This is possible because the priority order between student $i$
and $\emptyset $ for school $s$ may depend on the number of $\tau \left(
i\right) $-type students assigned to $s$. Moreover, we can consider a quota
restriction for a type. For $i$, $\emptyset \succ _{s}^{J}i$ if $\left\vert
J\cap I^{\tau \left( i\right) }\right\vert +1>\rho _{\tau \left( i\right) }$%
, where $\rho _{\tau \left( i\right) }$ represents the quota level for type $%
\tau \left( i\right) $. We formally discuss such a rule in Section 5.

Second, \textbf{WO} implies that\ the priority order between two students of
the same type also does not depend on the type distribution of the present
assignment. If \textbf{WO} is satisfied with $\succsim _{s}$ and $\tau
\left( i\right) =\tau \left( j\right) $, we simply write $i\succsim _{s}j$ ($%
j\succ _{s}i$ and $i\sim _{s}^{J}k$), by omitting $J$.

Third, we explain \textbf{DC}. This requires that the priority order between
two students be affected only by type distribution.\ If $\tau ^{s}\left(
i\right) =\tau ^{s}\left( j\right) $, then $J\cup \left\{ i\right\} $ and $%
J\cup \left\{ j\right\} $ have the same type distribution and thus the
priority order with $J\cup \left\{ i\right\} $ and that with $J\cup \left\{
j\right\} $ are essentially the same.\ To better understand \textbf{DC},
suppose that $\succsim _{s}$ satisfies \textbf{PD }and\textbf{\ WO}.
Moreover, additionally assume that there is $i^{\prime }\notin J$ such that $%
\tau ^{s}\left( i^{\prime }\right) =\tau ^{s}\left( i\right) =\tau
^{s}\left( j\right) $. Then, by \textbf{PD}, $j\succ _{s}^{J\cup \left\{
i\right\} }k$ implies $j\succ _{s}^{J\cup \left\{ i^{\prime }\right\} }k$.
Then, by transitivity, $i\succsim _{s}j$\ and $j\succ _{s}^{J\cup \left\{
i^{\prime }\right\} }k$ implies $i\succ _{s}^{J\cup \left\{ i^{\prime
}\right\} }k$. Then, again, by \textbf{PD}, $i\succ _{s}^{J\cup \left\{
j\right\} }k$. We similarly show that if $k\succ _{s}^{J\cup \left\{
j\right\} }i$, then $k\succ _{s}^{J\cup \left\{ i\right\} }j$. Therefore, if
such a student $i^{\prime }$ exists, then \textbf{PD }and\textbf{\ WO} imply 
\textbf{DC}. We require this to be satisfied even if there is no such a
student $i^{\prime }$.

Note that, \textbf{PD }and\textbf{\ WO} alone are insufficient for the
existence of stable matchings; that is, \textbf{DC} is also crucial. We show
this fact by introducing the following example.

\paragraph*{Example 1}

Let $I=\left\{ 1,2,3\right\} $ and $S=\{s\}$. The priority order profile $%
\succsim _{s}$ is assumed to satisfy 
\begin{equation*}
1\succ _{s}^{\{3\}}2\succ _{s}^{\left\{ 3\right\} }\emptyset ,\text{ }3\succ
_{s}^{\{2\}}1\succ _{s}^{\left\{ 2\right\} }\emptyset ,\text{ }2\succ
_{s}^{\{1\}}3\succ _{s}^{\left\{ 1\right\} }\emptyset .
\end{equation*}%
If two of the students are of the same type and the other is of a different
type from the two, then $\succsim _{s}$ satisfies \textbf{PD }and\textbf{\ WO%
}, but does not satisfy \textbf{DC}.\footnote{%
Further, the followings are necessary for \textbf{PD }and\textbf{\ WO}. If $%
\tau \left( 1\right) =\tau \left( 2\right) $, then $1\succ _{s}^{\emptyset
}2\succ _{s}^{\emptyset }3$. If $\tau \left( 2\right) =\tau \left( 3\right) $%
, then $2\succ _{s}^{\emptyset }3\succ _{s}^{\emptyset }1$. If $\tau \left(
3\right) =\tau \left( 1\right) $, then $3\succ _{s}^{\emptyset }1\succ
_{s}^{\emptyset }2$.} Suppose $\tau \left( 1\right) =\tau \left( 2\right) $
and $1\succ _{s}^{\emptyset }2\succ _{s}^{\emptyset }3$. Then, since $%
1\succsim _{s}^{\left\{ 3\right\} }2$ and $2\succ _{s}^{\left\{ 1\right\} }3$%
, but $3\succsim _{s}^{\left\{ 2\right\} }1$, $\succsim _{s}$ does not
satisfy \textbf{DC}.

Further, suppose $sP_{i}\emptyset $ for each $i=1,2,3$ and $q_{s}=2$. Any
non-wasteful matching $\mu $ satisfies $\left\vert \mu \left( s\right)
\right\vert =2$. Then, $3$ has justified envy toward $1$ at $\mu \left(
s\right) =\left\{ 1,2\right\} $, $1$ has justified envy toward $2$ at $\mu
^{\prime }\left( s\right) =\left\{ 2,3\right\} $ and $2$ has justified envy
toward $3$ at $\mu ^{\prime \prime }\left( s\right) =\left\{ 1,3\right\} $.
Hence, there is no stable matching in this example.\newline

On the other hand, as shown later, if $\succsim _{s}$ satisfies \textbf{PD},%
\textbf{\ WO, }and\textbf{\ DC}, then there must exist stable matchings. The
following two results are the keys of the fact.

First, to show our results, the following two technical results are
important.

\begin{lemma}
Suppose that $\succsim _{s}$ satisfies \textbf{PD},\textbf{\ WO, }and\textbf{%
\ DC}. Then, for any $J\in \mathcal{J}_{s}$ and any $i,j,k\in \left(
I\setminus J\right) $, $i\succ _{s}^{J\cup \left\{ k\right\} }j$ and $j\succ
_{s}^{J\cup \left\{ i\right\} }k$ imply $i\succ _{s}^{J\cup \left\{
j\right\} }k$.
\end{lemma}

The proof of this result is in Appendix A.

This result implies that $\succsim _{s}$ satisfies a kind of transitivity
even if the given assignments slightly differ.

Next, to facilitate later explanations, we introduce specific terminology
regarding $\succsim _{s}$. For a given $J\subseteq I$ with $\left\vert
J\right\vert \in \left[ 2,q_{s}+1\right] $, $i\in J$ has the \textbf{lowest
(highest) priority} \textbf{within} $J$ if there is no $j\in J$ such that $j$
has a lower priority\textbf{\ }than $i$ with $J\setminus \left\{ i,j\right\} 
$. We show that with our assumptions for any $J\subseteq I$ with $\left\vert
J\right\vert \in \left[ 2,q_{s}+1\right] $, there is a student in $J$ who
has the lowest priority within $J$.

\begin{lemma}
\textit{Suppose that }$\succsim _{s}$\textit{\ satisfies \textbf{PD},\ 
\textbf{WO}},\textit{\ and\ \textbf{DC}. Then, for any }$J$\textit{\ such
that }$\left\vert J\right\vert \in \left[ 2,q_{s}+1\right] $,\textit{\ there
is }$i\in J$\textit{\ that has the lowest priority within }$J$\textit{; that
is, }$j\succsim _{s}^{J\setminus \left\{ i,j\right\} }i$\textit{\ for all }$%
j\in J\setminus \left\{ i\right\} $\textit{.}
\end{lemma}

The proof of this result is in Appendix A.

This result implies that for any $J$, we can determine the student who has
the lowest priority within\textit{\ }$J$. Thus, in the SPDA-like mechanism
introduced later, a rejected student can be consistently chosen from $J$.

Hereafter, we assume that $\succsim _{s}$ satisfies \textbf{PD, WO, }and%
\textbf{\ DC}. In Section 5, we introduce a class of priority rules called
adjusted scoring rules, such that if $\succsim _{s}$ satisfies \textbf{PD, WO%
},\textbf{\ }and\textbf{\ DC}. This class includes the\ several rules
introduced in previous studies such as those with type quotas and/or
reserves. Moreover, with this class of priority rules, we can take into
account a trade-off explained in Section 1 more flexibly.

\section{Mechanisms}

For a given $\succsim $, let $\phi :\mathcal{P}^{\left\vert I\right\vert
}\rightarrow \mathcal{M}$ be a mechanism, where $\mathcal{M}$ is the set of
possible matchings. We introduce three properties on mechanisms.\ First, $%
\phi $ is a \textbf{stable mechanism}\textit{\ }if $\phi \left( P\right) $
is stable for any $P\in \mathcal{P}^{\left\vert I\right\vert }$. Second, $%
\phi $ is a \textbf{student-optimal stable mechanism} if for any $P\in 
\mathcal{P}^{\left\vert I\right\vert }$, $\phi \left( P\right) $ is stable
and $\phi (P)$\ is not Pareto dominated by any other stable matchings (for $%
P $). Finally, $\phi $ is a \textbf{group strategy-proof mechanism} if for
any $P\in \mathcal{P}^{\left\vert I\right\vert }$, there is no group $%
J\subseteq I$ and $P_{J}^{\prime }\in \mathcal{P}^{\left\vert J\right\vert }$
such that%
\begin{equation*}
\phi \left( P_{J}^{\prime },P_{-J}\right) \left( j\right) P_{j}\phi \left(
P\right) \left( j\right) ,
\end{equation*}%
for all $j\in J$.

Our mechanism is constructed as follows. First, a choice function of each
school $s\in S$ is derived from $\succsim _{s}$. Second, a matching is
derived by the Gale and Shapley's (1962) SPDA algorithm with the choice
functions. Then, by utilizing two results introduced by Roth and Sotomayor
(1990) and Hatfield and Kojima (2009), we can show that this mechanism is
stable and strategy-proof.

We introduce the choice function only for such applications. Let $%
C_{s}:2^{I}\rightarrow 2^{I}$ be a choice function of $s$ satisfying $%
C_{s}\left( J\right) \subseteq J,$ $\left\vert C_{s}\left( J\right)
\right\vert \leq q_{s},$ and $C=\left( C_{s}\right) _{s\in S}$. Moreover,
let $\phi _{DA}^{C}$ be the SPDA mechanism with respect to $C$. The formal
definition of the SPDA algorithm with $C$ is provided in Appendix B.

We introduce a method to construct a choice function of $s$ from $\succsim
_{s}$ denoted by $C_{s}^{\succsim }$. First, without loss of generality, we
let $T=\{0,1,\ldots ,\left\vert T\right\vert -1\}$ with $\tau \left(
\emptyset \right) =0$ and $I=\{1,2,\ldots ,\left\vert I\right\vert \}$ such
that, for $i,j\in I$,

\begin{description}
\item[(i)] if $\tau \left( i\right) <\tau \left( j\right) $, then $i<j$,

\item[(ii)] if $\tau \left( i\right) =\tau \left( j\right) $ and $i<j$, then 
$i\succsim _{s}j$.
\end{description}

In words, first, if the types of two students differ, then the student of
the smaller type number is assigned to a smaller number than that assigned
to the other. Second, if the two students have the same type, then the
student assigned to the larger number never takes priority over the other.
This construction is possible, because of \textbf{WO}.

Fix $J\subseteq I$. Let $J=\left\{ a_{1},\ldots ,a_{\left\vert J\right\vert
}\right\} $ satisfying $a_{1}<a_{2}<\ldots <a_{\left\vert J\right\vert }$.

\textbf{Step} $0$ Let $C_{s}^{0}\left( J\right) =\emptyset $

\textbf{Step} $g\in \left[ 1,\left\vert J\right\vert \right] $ First, if $%
\left\vert C_{s}^{g-1}\left( J\right) \right\vert <q_{s}$ and $a_{g}\succsim
_{s}^{C_{s}^{g-1}\left( J\right) }\emptyset $, then $C_{s}^{g}\left(
J\right) =C_{s}^{g-1}\left( J\right) \cup \left\{ a_{g}\right\} $ and go to
Step $g+1$. Second, if $\left\vert C_{s}^{g-1}\left( J\right) \right\vert
<q_{s}$ and $\emptyset \succ _{s}^{C_{s}^{g-1}\left( J\right) }a_{g}$, then $%
C_{s}^{g}\left( J\right) =C_{s}^{g-1}\left( J\right) $ and go to Step $g+1$.
Third, if $\left\vert C_{s}^{g-1}\left( J\right) \right\vert =q_{s}$, then
we choose $r_{g}$ who most recently applies to $s$ among the students who
have the lowest priority within $C_{s}^{g-1}\left( J\right) \cup \left\{
a_{g}\right\} $. Then, $C_{s}^{g}\left( J\right) =C_{s}^{g-1}\left( J\right)
\cup \left\{ a_{g}\right\} \setminus \left\{ r_{g}\right\} $ and go to Step $%
g+1$.

Finally, let $C_{s}^{\succsim }\left( J\right) =C_{s}^{\left\vert
J\right\vert }\left( J\right) $ for all $J$ and $s$, and $C^{\succsim
}=\left( C_{s}^{\succsim }\right) _{s\in S}$.\newline

Note that the steps above can be interpreted as being similar to the SPDA,
as students in $J$ appear to apply to $s$ one by one, and $s$ seems to
accept one student while rejecting another in each step. However, this
process merely constructs a choice function. Therefore, we run the SPDA
separately under $C^{\succsim }$. That is, we let $\bar{\phi}$ be a
mechanism such that 
\begin{equation*}
\bar{\phi}\left( P\right) \left( s\right) =\phi _{DA}^{C^{\succsim }}\left(
P\right) \left( s\right)
\end{equation*}%
for all $s$ and all $P\in \mathcal{P}^{\left\vert I\right\vert }$, where $%
C^{\succsim }=\left( C_{s}^{\succsim }\right) _{s\in S}$.

\begin{theorem}
$\bar{\phi}$ is a stable and group strategy-proof mechanism.
\end{theorem}

The proof of this result is provided in Appendix B. Briefly, the proof
proceeds as follows. First, we show if $\mu $ is stable under $C^{\succsim }$%
, then $\mu $ is also stable under $\succsim $. Second, the fact $\phi
_{DA}^{C^{\succsim }}\left( P\right) $ is stable under $C^{\succsim }$
follows from the substituteness of $C_{s}^{\succsim }$ for all $s$ and a
classical result of Roth and Sotomayor (1990). Consequently, $\bar{\phi}%
\left( P\right) $ is stable (under $\succsim $). The strategy-proofness of $%
\bar{\phi}$ follows from the fact that $C_{s}^{\succsim }$ is substitutable
and size monotonic, which is known to be sufficient as shown by Hatfield and
Kojima (2009). In Appendix B, we show that $C_{s}^{\succsim }$ is
substitutable and size monotonic.

We provide an intuitive explanation of why $\bar{\phi}$ is stable and group
strategy-proof. First, we explain the fairness of $\bar{\phi}\left( P\right) 
$. In rounds of the SPDA with $C^{\succsim }$, a student who has the lowest
priority within the set of the temporarily matched students and the
applicant is rejected. By Lemma 2, it is feasible to choose such a student
while satisfying \textbf{PD, WO, }and\textbf{\ DC}. Let $r$ be a student who
is rejected by a school $s$ in a round of the SPDA. We consider a possible
scenario in which $r$ has justified envy toward student $i$, who is
temporarily matched to $s$ after that round.

This situation can arise only if an applicant $a$ of the same type as $i$ is
temporarily accepted in a round after the one in which $r$ was rejected. Let 
$r^{\prime }$ be the student who is rejected by $s$ in that round. By 
\textbf{PD}, $\tau \left( a\right) \neq \tau \left( r^{\prime }\right) $,
because otherwise, the type distribution remains unchanged. Since $r^{\prime
}$ is rejected while $i$ is not, $i$ does not have a lower priority than $%
r^{\prime }$ with the set of new temporarily accepted students. Moreover, by 
\textbf{WO}, this situation can occur only when $\tau \left( r\right) \neq
\tau \left( a\right) =\tau \left( i\right) $, since $i\succsim _{s}r$ holds
for any set of students in the case where $\tau \left( r\right) =\tau \left(
i\right) $.

Hence, the numbers of $\tau \left( r^{\prime }\right) $-type students and $%
\tau \left( r\right) $-type students remain unchanged. Since $r^{\prime }$
does not have a lower priority than $r$ with the set of previous temporarily
accepted students, $r^{\prime }$ also does not have a lower priority than $r$
with the set of new temporarily accepted students. Thus, by transitivity, we
conclude that $r$ does not have a higher priority than $i$ with the set of
new temporarily accepted students. Therefore, it is impossible that $r$ has
justified envy toward\ any students in $\phi _{DA}^{C^{\succsim }}\left(
P\right) \left( s\right) $.

This implies that an SPDA result will be stable if in each round, a student
with the lowest priority for $s$ within the set of the temporarily matched
students and the applicant to school $s$ is rejected by $s$. In this study,
the priority order for $s$ with $J\in \mathcal{J}_{s}$ is weak order. Thus,
there may exist multiple students have the lowest priority in $s$. This
implies that a tie-breaking rule to determine one rejected student among the
students who have the lowest priority is necessary.

If the tie-breaking rule is inappropriate, a student may have an incentive
to manipulate their preferences, as applying early or late could be
advantageous due to the tie-breaking rule. However, under the rule that
determines the rejected students through $C_{s}^{\succsim }$, no student has
such incentives, because the timing of their application does not affect
their chances of being accepted under $C_{s}^{\succsim }$. Consequently, $%
\bar{\phi}$ is strategy-proof.

Since this mechanism works even when $\succsim _{s}^{J}$ is not a linear
order but a weak order for all $J\in \mathcal{J}_{s}$ and all $s$, we
succeed in extending the results of several previous studies to accommodate
cases in which priorities may include ties. Section 5 provides further
discussion on this point. However, $\bar{\phi}$ may not be student-optimally
stable. We show this fact by providing the following example.

\paragraph*{Example 2}

Let $I=\left\{ 1,2,3,4\right\} $ and $S=\{s,s^{\prime }\}$. We assume $%
q_{s}=2$ and $q_{s^{\prime }}=1$. Moreover, let

\begin{equation*}
\tau \left( 1\right) =\tau \left( 2\right) =1\text{, and }\tau \left(
3\right) =\tau \left( 4\right) =2\text{.}
\end{equation*}%
The preferences of the students are 
\begin{equation*}
P_{1}:\text{ }6\text{ }5\text{ }\emptyset ,\text{ }P_{2}:\text{ }6\text{ }5%
\text{ }\emptyset ,\text{ }P_{3}:\text{ }5\text{ }6\text{ }\emptyset ,\text{ 
}P_{4}:\text{ }5\text{ }6\text{ }\emptyset .
\end{equation*}

The priority order profile $\succsim $ is assumed to satisfy 
\begin{gather*}
3\succ _{s}^{\{1\}}4\succ _{s}^{\left\{ 1\right\} }2\succ _{s}^{\left\{
1\right\} }\emptyset ,\text{ }3\succ _{s}^{\{2\}}4\succ _{s}^{\left\{
2\right\} }1\succ _{s}^{\left\{ 2\right\} }\emptyset , \\
2\succ _{s}^{\{3\}}1\sim _{s}^{\left\{ 3\right\} }4\succ _{s}^{\left\{
3\right\} }\emptyset ,\text{ }2\succ _{s}^{\{4\}}3\succ _{s}^{\{4\}}1\succ
_{s}^{\left\{ 4\right\} }\emptyset , \\
i\succ _{s}^{\emptyset }\emptyset \,\text{for all }i=1,2,3,4, \\
4\succ _{s^{\prime }}^{\emptyset }3\succ _{s^{\prime }}^{\emptyset }2\succ
_{s^{\prime }}^{\emptyset }1\succ _{s^{\prime }}^{\emptyset }\emptyset ,
\end{gather*}%
Then, by the construction above, the choice function of $s$ satisfies%
\begin{gather*}
C_{s}^{\succsim }\left( \left\{ 1,2,3,4\right\} \right) =\left\{ 2,3\right\}
,C_{s}^{\succsim }\left( \left\{ 1,2,3\right\} \right) =\left\{ 2,3\right\}
,C_{s}^{\succsim }\left( \left\{ 1,2,4\right\} \right) =\left\{ 2,4\right\}
\\
C_{s}^{\succsim }\left( \left\{ 1,3,4\right\} \right) =\left\{ 1,3\right\}
,C_{s}^{\succsim }\left( \left\{ 2,3,4\right\} \right) =\left\{ 2,3\right\} 
\text{.}
\end{gather*}%
For example, we consider the construction of $C_{s}^{\succsim }\left(
\left\{ 1,3,4\right\} \right) $. In this case, $a_{1}=1,$ $a_{2}=3$ and $%
a_{3}=4$. Then, $C_{s}^{1}\left( \left\{ 1,3,4\right\} \right) =\left\{
1\right\} ,$ $C_{s}^{2}\left( \left\{ 1,3,4\right\} \right) =\left\{
1,3\right\} $ and $C_{s}^{3}\left( \left\{ 1,3,4\right\} \right) =\left\{
1,3\right\} $, because $1$ and $4$ have the lowest priority within $\left\{
1,3,4\right\} $ and $4$ more recently applies to $s$.

Then, 
\begin{equation*}
\bar{\phi}\left( P\right) \left( s\right) =\left\{ 2,3\right\} \text{ and }%
\bar{\phi}\left( P\right) \left( s^{\prime }\right) =\left\{ 4\right\} \text{
(}1\text{ is unmatched).}
\end{equation*}%
This matching is stable but not student-optimally stable, because a matching 
$\mu $ such that$\ \mu (s)=\left\{ 3,4\right\} $ and $\mu (s^{\prime
})=\left\{ 1\right\} $ ($1$ is unmatched) Pareto dominates $\bar{\phi}\left(
P\right) $ and is also stable.\newline
Therefore, for some $\succsim $, $\bar{\phi}\left( P\right) $ is not
student-optimal stable mechanism. However, in the following specific case,
then $\bar{\phi}$ is a student-optimal stable mechanism.

\begin{theorem}
If $\succsim _{s}^{J}$ is a linear order for all $J\in \mathcal{J}_{s}$ and
all $s$, then $\bar{\phi}$ is a student-optimal stable and group
strategy-proof mechanism.
\end{theorem}

The proof of this result is provided in Appendix B.

Thus, if $\succsim _{s}^{J}$ is a linear order for all $J\in \mathcal{J}_{s}$
and all $s$, then we can avoid the stable matchings that are Pareto
dominated by another stable matching. Since many of the previous studies
focus on the case where priorities do not include any ties, Theorem 2 is a
generalization of their results. For example, this is a generalization of
Ehlers et al. (2014, Theorems 4 and 5), because as is shown in the
subsequent section and Appendix C, $\bar{\phi}\left( P\right) $ attains a
student-optimal stable matching in their definition.

We note the running time of the mechanism $\bar{\phi}$. Constructing $%
C_{s}^{\succsim }\left( J\right) $ for all $s\in S$ and all $J\subseteq I$
cannot be completed in polynomial time, because there are $2^{\left\vert
I\right\vert }$ possible subsets of $I$. Thus, to computationally
efficiently execute $\bar{\phi}$, we derive $C_{s}^{\succsim }\left(
J\right) $ only as needed. In Appendix D, we show that $\bar{\phi}$ can be
executed efficiently.

\section{Adjusted scoring rule}

In this section, we provide a subclass of priority order profiles satisfying 
\textbf{PD, WO, }and \textbf{DC}, which not only covers the priority order
profiles corresponding to the priority (or choice) rules in the previous
studies, but also can directly reflect more subtle trade-offs as discussed
in Section 1.

For $s\in S,$ let $\sigma _{i}^{s}\in \left[ 0,1\right] $ for each $i\in I$
and $\alpha _{t}^{s}:\mathbb{Z}_{++}\rightarrow \mathbb{R}$ be a
non-increasing function for all $t\in T$. Let $\sigma ^{s}=\left( \sigma
_{i}^{s}\right) _{i\in I}$ and $\alpha ^{s}=\left( \alpha _{t}^{s}\right)
_{t\in T}$. For example, $\sigma _{i}^{s}$ represents the test score of $i$
for the entrance exam of school $s$. Moreover, let $A$ be the set of
possible non-increasing functions; that is, $\alpha ^{s}\in A$. Since we
consider only one school to explain this rule here, we omit $s$ from $\sigma
^{s}$ and $\alpha ^{s}$ for notational simplicity.

First, we assume the following rule: Let $f_{i}(J)=\sigma _{i}+\alpha _{\tau
\left( i\right) }\left( \left\vert J\cap I^{\tau \left( i\right)
}\right\vert \right) $ for $i\in I$ representing an adjusted score.
Moreover, $f_{\emptyset }(J)=\underline{\sigma }$, which is the threshold of
the adjusted score of an acceptable student. We say that $\succsim _{s}$ is 
\textbf{an adjusted scoring rule with respect to} $\left( \sigma ,\underline{%
\sigma },\alpha \right) \in \left[ 0,1\right] ^{I}\times \mathbb{R}\times A$
if for all $J\in \mathcal{J}_{s}$ and all $i,j\in \left( I\cup \left\{
\emptyset \right\} \right) \setminus J$, $i\succ _{s}^{J}j$ if $%
f_{i}(J)>f_{j}(J)$ and $i\sim _{s}^{J}j$ if $f_{i}(J)=f_{j}(J)$.

\begin{proposition}
If $\succsim _{s}$ is an adjusted scoring rule with respect to $\left(
\sigma ,\underline{\sigma },\alpha \right) \in \left[ 0,1\right] ^{I}\times 
\mathbb{R}_{+}\times A$, then $\succsim _{s}$ satisfies \textbf{PD, WO},%
\textbf{\ }and \textbf{DC}.
\end{proposition}

Since the proof of this result is trivial, we omit it here.

This class of rules is a generalization of those of several previous studies
on matching with diversity constraints such as Ehlers et al. (2014),
Echenique and Yenmez (2015), Erdil and Kumano (2019) and Imamura (2023). We
explain this below.

Let $r=\left( r_{t}\right) _{t\in T}\in \mathbb{Z}_{+}^{_{\left\vert
T\right\vert }}$.\footnote{%
Note that $r$ is allowed to differ across different schools.} Moreover, let%
\begin{equation*}
\alpha _{\tau \left( i\right) }\left( x\right) =2\left( r_{\tau \left(
i\right) }-\left( x+1\right) \right)
\end{equation*}%
for all $i\in I$, all $s\in S$\ and all $x\in \mathbb{Z}_{++}$. Then, $%
i\succ _{s}^{J}j$ if and only if either $r_{\tau \left( i\right)
}-\left\vert J\cap I^{\tau \left( i\right) }\right\vert >r_{\tau \left(
j\right) }-\left\vert J\cap I^{\tau \left( j\right) }\right\vert $, or $%
r_{\tau \left( i\right) }-\left\vert J\cap I^{\tau \left( i\right)
}\right\vert =r_{\tau \left( j\right) }-\left\vert J\cap I^{\tau \left(
j\right) }\right\vert $ and $\sigma _{i}>\sigma _{j}$. Moreover, let $%
\underline{\sigma }=-\infty $. In this case, the model is (effectively)
equivalent to that discussed by Erdil and Kumano (2019). Kitahara and
Okumura (2023a) formally show the equivalence. In Appendix C, we provide $%
\left( \sigma ,\underline{\sigma },\alpha \right) ,$ in which $C^{\succsim }$
is equivalent to those introduced by Echenique and Yenmez (2015) and Imamura
(2023). Moreover, in Appendix C, we also show that for some $\left( \sigma ,%
\underline{\sigma },\alpha \right) $, our model is (effectively) equivalent
to that discussed by Ehlers et al. (2014).

Next, we consider the priority-based rule discussed by several studies
introduced in Section 2 and actually used in China. According to Wang
(2009), in Chinese colleges, admission success or failure is determined by
examination scores, and in several provinces/regions, bonus points are given
to minority applicants.\footnote{%
The University of Michigan in the United States also had such a rule,
although it is no longer used. See Gratz v. Bollinger, 539 U.S. 244:
https://supreme.justia.com/cases/federal/us/539/244/} For example, in
Xinjang, there are three types of applicants. The applicants of the first
type (say type $A$) are those with both parents belonging to the ethnic
minority group. The applicants of the second type (say type $B$) are those
with one parent belonging to the group. The remaining applicants belong to
the third type (say type $C$). Then, in the rule of the region, $\alpha
_{A}\left( x\right) =50,$ $\alpha _{B}\left( x\right) =10$ and $\alpha
_{C}\left( x\right) =0$ for all $x\in \mathbb{Z}_{++}$. However, in the
adjusted scoring rule, the bonus points for minority students at a school
can be dependent on the type distribution of the school.

Thus far, we explain that our rule includes the rules in previous studies
and those in the real-world. Next, we discuss the advantages that are not
present in other rules of ours, by introducing the example below.

\paragraph*{Example 3}

Suppose that the students are divided into two types $A$ and $B$, which
represents advantaged and disadvantaged students, respectively. Let $J$ be
the set of applicants of school $s$. The objective function of school $s$
denoted by $v_{s}\left( J^{\prime }\right) $ is assumed to be as follows: 
\begin{equation}
v_{s}\left( J^{\prime }\right) =\left\vert J^{\prime }\right\vert
-\sum\nolimits_{j\in J^{\prime }}\left( 1-\sigma _{j}\right) +\frac{1}{2}%
\sqrt{\left\vert J^{\prime }\cap I^{B}\right\vert },  \label{c}
\end{equation}%
where $J^{\prime }$ represents a set of students and $\sigma _{j}$
represents the test score of student $j$. The first term of (\ref{c})
represents the benefit of assigning students regardless of their types. For
the second term of (\ref{c}), $1-\sigma _{j}$ is the cost for assigning
student $j$; that is, the associated cost for a student will be large when
her score is low. The third term of (\ref{c}) represents the benefits from
assigning disadvantaged students. From the viewpoint of diversity, the
benefit is a convex function of the number of disadvantaged students. School 
$s$ decides $J^{\ast }$ to maximize (\ref{c}) subject to $J^{\ast }\subseteq
J$ and $\left\vert J^{\ast }\right\vert \leq q_{s}=5$. Thus, the school
prefers $i$ over $j$ for adding to $J^{\prime }$ if and only if $v\left(
J^{\prime }\cup \left\{ i\right\} \right) >v\left( J^{\prime }\cup \left\{
j\right\} \right) $. If $\tau \left( i\right) =B$ and $\tau \left( j\right)
=A$, then the condition is 
\begin{equation}
\sigma _{i}-\sigma _{j}+\frac{1}{2}\left( \sqrt{\left\vert J^{\prime }\cap
I^{B}\right\vert +1}-\sqrt{\left\vert J^{\prime }\cap I^{B}\right\vert }%
\right) >0.  \label{d}
\end{equation}

We assume that $J=\left\{ 1,\dots,6\right\} $ described in the following
table.

\begin{center}
\begin{tabular}{|l|l|l|l|l|l|l|}
\hline
Student $i$ & 1 & 2 & 3 & 4 & 5 & 6 \\ \hline
Type $\tau \left( i\right) $ & $A$ & $A$ & $A$ & $A$ & $B$ & $B$ \\ \hline
Score $\sigma _{i}$ & 0.8 & 0.8 & 0.7 & 0.6 & 0.5 & $\sigma _{6}$ \\ \hline
\end{tabular}
\end{center}

If $\sigma _{6}>0.41$, then $J^{\ast }=\{1,2,3,5,6\}$. However, if $\sigma
_{6}<0.39$, then $J^{\ast }=\{1,2,3,4,5\}$. To achieve $J^{\ast }$ by using
the reserves rule, the optimal number of reserved seats for type $B$ denoted
by $r_{B}^{\ast }$ is dependent on $\sigma _{6}$. That is, $r_{B}^{\ast }=2$
if $\sigma _{6}>0.4$ and $r_{B}^{\ast }=1$ if $\sigma _{6}<0.39$. Since a
school must often decide the number before observing the scores of the
applicants, the school makes a decision based on an uncertain expectation of
the scores.

Next, we consider the priority-based rule. In the case where $1,\ldots ,6$
are the applicants, then we can achieve $J^{\ast }$ via the priority-based
rule by awarding about $0.2$ bonus point to the type $B$ applicants.
However, for example, let $3^{\prime }$ be a type $B$ student whose score is 
$\sigma _{3^{\prime }}=0.7$.$\,$If $3^{\prime }$ applies to $s$ instead of $%
3 $ and $0.4<\sigma _{6}<0.44$, then the bonus point is excessive. This is
because, in that case, three type $B$ students $3^{\prime },$ $5$ and $6$
are chosen and $4$ is not, but $J^{\ast }=\{1,2,3^{\prime },4,5\}$. Thus,
the optimal bonus points of a school also depend on the students who apply
to that school.

However, by letting 
\begin{equation}
\alpha _{A}\left( x\right) =0,\text{ }\alpha _{B}\left( x\right) =\frac{1}{2}%
\left( \sqrt{x+1}-\sqrt{x}\right)  \label{w}
\end{equation}%
for all $x\geq 0$, we have $i\succ _{s}^{J^{\prime }}j$ if and only if $%
v\left( J^{\prime }\cup \left\{ i\right\} \right) >v\left( J^{\prime }\cup
\left\{ j\right\} \right) $, because of (\ref{d}). That is, by using the
adjusted scoring rule with (\ref{w}), school $s$ can always achieve the
optimal subset of the applicants.\newline

Finally, S\"{o}nmez and Switzer (2013) and Kominers and S\"{o}nmez (2016)
discuss models with slot-specific priorities, where each school has a number
of slots and each of them has a total order priority. Since the choice of
school is sequentially determined one by one in our construction of $%
C_{s}^{\succsim }$, our approach seems similar to the slot-specific priority
one. However, in as shown in Appendix E, for the objective function given by
(\ref{c}), the optimal choice may not be achieved for any slot-specific
priorities.

\section{Concluding Remarks}

We provide a strategy-proof mechanism based on the SPDA that can be used to
derive a stable matching in cases in which the priority orders for schools
respect diversity constraints and are weak orders with each given
assignment. Moreover, if the priority order for each school is restricted to
be a linear order with each given assignment, then the mechanism is
student-optimal stable.

Regarding the limitations of this study, Kurata et al. (2017) and S\"{o}nmez
and Yenmez (2022) note that in real-life applications, a student may belong
to multiple types. For example, students may be categorized by gender (male
or female) as well as by ethnic background (African, Asian, European, Latin
American, etc.). In our analysis, addressing this situation would require
defining eight or more types (e.g., male African students, female African
students, and so on). However, our framework cannot fully incorporate
priority dependence under multiple types. Specifically, the priority order
between two students of different types depends solely on the number of
students in each type due to \textbf{PD}. However, it may be more
appropriate for the priority order between these students to also depend on
the number of students in another type.

We illustrate this point by considering two extreme school assignments. In
one scenario, all assigned students are Asian women, and in the other, they
are all Asian men. We assume one seat is still available in both
assignments. Now, consider two applicants: a female African student and a
male European student. To enhance diversity, we would prefer choosing the
female African student in the former assignment and the male European
student in the latter assignment. However, \textbf{PD} requires that the
priority between the two to be the same in both assignments. Therefore, our
framework may not be suitable in such situations.\footnote{%
The frameworks of Westkamp (2013) and Ayg\"{u}n and Turhan (2020) may be
better to engage in this problem, because they allow an increase in the
number of students of a type to change the priority between students with
other two types.}

In this study, we consider priority orders for schools as weak orders with
each given assignment. In a recent study, Kitahara and Okumura (2021)
examine a case where school priority orders are not weak orders but partial
orders. Unfortunately, as shown by Kitahara and Okumura (2023c) if school
priority orders are not weak orders with some assignments, the mechanism
introduced in this study fails to ensure stability. To address this issue,
the previous version introduced an alternative mechanism that remains stable
even in such cases. However, that mechanism is not strategy-proof, even when
school priority orders are weak orders with each given assignment.

\section*{References}

\begin{description}
\item Abdulkadiro\u{g}lu, A. 2005. \textquotedblleft College admissions with
affirmative action,\textquotedblright\ International Journal of Game Theory
33, 535--549.

\item Abdulkadiro\u{g}lu, A., S\"{o}nmez, T. 2003. \textquotedblleft School
choice: A mechanism design approach,\textquotedblright\ American Economic
Review 93(3), 729--747.

\item Arcidiacono, P., Aucejo, E.M., Hotz, V.J. 2016. \textquotedblleft
University Differences in the Graduation of Minorities in STEM Fields:
Evidence from California.\textquotedblright\ American Economic Review 106
(3), 525--562.

\item Arcidiacono, P., Aucejo, E.M., Kenneth, H.F., Spenner, K.I., 2011.
\textquotedblleft Does affirmative action lead to mismatch? A new test and
evidence,\textquotedblright\ Quantitative Economics 2(3), 303-333.

\item Arcidiacono, P., Lovenheim, M. 2016. \textquotedblleft Affirmative
action and the quality-fit trade-off,\textquotedblright\ Journal of Economic
Literature 54(1), 3-51.

\item Ayg\"{u}n. O., Turhan, B. 2020. \textquotedblleft Dynamic reserves in
matching markets,\textquotedblright\ Journal of Economic Theory 188, 105069.

\item Dur, U., Pathak, P. A., S\"{o}nmez, T., 2020. \textquotedblleft
Explicit vs. statistical targeting in affirmative action: Theory and
evidence from Chicago's exam schools,\textquotedblright\ Journal of Economic
Theory 187, 104996.

\item Echenique, F., Yenmez, M.B., 2015. \textquotedblleft How to control
controlled school choice,\textquotedblright\ American Economic Review 105,
2679--2694.

\item Ehlers, L., Hafalir, I.E., Yenmez, M.B., Yildirim, M.A. 2014.
\textquotedblleft School Choice with Controlled Choice Constraints: Hard
Bounds versus Soft Bounds,\textquotedblright\ Journal of Economic Theory
153, 648--683.

\item Erdil, A., Ergin, H. 2008. \textquotedblleft What's the matter with
tie-breaking? Improving efficiency in school choice,\textquotedblright\
American Economic Review 98(3), 669--689.

\item Erdil, A., Kumano, T. 2019. \textquotedblleft Efficiency and stability
under substitutable priorities with ties,\textquotedblright\ Journal of
Economic Theory 184, 104950.

\item Gale, D., Shapley, L.S. 1962. \textquotedblleft College admissions and
the stability of marriage\textquotedblright , American Mathematical Monthly
69(1), 9-15.

\item Hafalir, I.E., Yenmez, M.B., Yildirim, M.A. 2013. Effective
Affirmative Action in School Choice,\ Theoretical Economics 8(2), 325--363.

\item Imamura, K. 2023. Meritocracy versus Diversity, Mimeo.

\item Kitahara, M., Okumura, Y. 2021. \textquotedblleft Improving Efficiency
in School Choice Under Partial Priorities,\textquotedblright\ International
Journal of Game Theory 50(4), 971-987

\item Kitahara, M., Okumura, Y. 2023a. \textquotedblleft Stable Improvement
Cycles in a Controlled School Choice,\textquotedblright\ Available at SSRN:
https://ssrn.com/abstract=3582421

\item Kitahara, M., Okumura, Y. 2023b. \textquotedblleft School Choice with
Multiple Priorities,\textquotedblright\ Available at arXiv:2308.04780

\item Kitahara, M, Okumura, Y. 2023c \textquotedblleft Stable Mechanisms in
Controlled School Choice,\textquotedblright\ Available at SSRN:
https://ssrn.com/abstract=4374711

\item Kojima, F. 2012. \textquotedblleft School Choice: Impossibilities for
Affirmative Action,\textquotedblright\ Games and Economic Behavior 75(2),
685--693

\item Kominers, S.D., S\"{o}nmez, T. 2016. \textquotedblleft Matching with
slot-specific priorities: Theory.\textquotedblright\ Theoretical Economics
11(2), 683-710.

\item Kurata, R., Hamada, N., Iwasaki, A. and Yokoo, M. 2017.
\textquotedblleft Controlled school choice with soft bounds and overlapping
types,\textquotedblright\ Journal of Artificial Intelligence Research 58,
153--184.

\item Roth, A.E., Sotomayor, M.A.O. 1990. Two-Sided Matching: A Study in
Game-theoretic Modeling and Analysis. Econometric Society Monographs.

\item Sander, R. H., Taylor Jr, S. 2012. Mismatch: How Affirmative Action
Hurts Students It's Intended to Help, and Why Universities Won't Admit It,
Basic Books.

\item S\"{o}nmez, T., Switzer, T.B. 2013. \textquotedblleft Matching with
(branch-of-choice) contracts at United States Military
Academy,\textquotedblright\ Econometrica, 81: 451-488.

\item S\"{o}nmez, T., Yenmez, M.B. 2022. \textquotedblleft Affirmative
action in India via vertical, horizontal, and overlapping
reservations,\textquotedblright\ Econometrica\ 90(3), 1143-1176.

\item Wang, T. 2009. \textquotedblleft Preferential policies for minority
college admission in China: Recent developments, necessity, and
impact,\textquotedblright\ In M. Zhou and A.M. Hill (eds) Affirmative action
in China and the U.S., Palgrave Macmillan.

\item Westkamp, A. 2013. \textquotedblleft An Analysis of the German
university admissions system,\textquotedblright\ Economic Theory 53, 561-589.
\end{description}

\newpage

\section*{Appendix A}

\subsection*{Proof of Lemma 1.}

First, suppose $\tau \left( i\right) =\tau \left( j\right) \neq \tau \left(
k\right) $. By \textbf{WO}, $i\succ _{s}^{J\cup \left\{ k\right\} }j$
implies $i\succ _{s}j$. By \textbf{DC}, $i\succ _{s}j$ and $j\succ
_{s}^{J\cup \left\{ i\right\} }k$\textbf{\ }imply $i\succ _{s}^{J\cup
\left\{ j\right\} }k$.

Second, suppose $\tau \left( i\right) \neq \tau \left( j\right) =\tau \left(
k\right) $. Then, by \textbf{WO}, $j\succ _{s}^{J\cup \left\{ i\right\} }k$
implies $j\succ _{s}k$. By \textbf{DC, }$i\succ _{s}^{J\cup \left\{
k\right\} }j$ and $j\succ _{s}k$\textbf{\ }imply $i\succ _{s}^{J\cup \left\{
j\right\} }k$.

Third, suppose $\tau \left( i\right) =\tau \left( k\right) \neq \tau \left(
j\right) $. We prove this case by contraposition. Suppose $k\succsim _{s}i$.
We show that $j\succsim _{s}^{J\cup \left\{ k\right\} }i$ or $k\succsim
_{s}^{J\cup \left\{ i\right\} }j$. By \textbf{DC}, $i\succ _{s}^{J\cup
\left\{ k\right\} }j$ implies $k\succ _{s}^{J\cup \left\{ i\right\} }j$, and 
$j\succ _{s}^{J\cup \left\{ i\right\} }k$ implies $j\succ _{s}^{J\cup
\left\{ k\right\} }i$. Therefore, by contraposition, we have $i\succ
_{s}^{J\cup \left\{ k\right\} }j$ and $j\succ _{s}^{J\cup \left\{ i\right\}
}k$ imply $i\succ _{s}^{J\cup \left\{ j\right\} }k$.

Fourth, if $\tau \left( i\right) =\tau \left( j\right) =\tau \left( k\right) 
$, then by \textbf{WO}, $i\succ _{s}j$ and $j\succ _{s}k$, which by
transitivity imply $i\succ _{s}k$.

Finally, suppose $\tau \left( i\right) ,\tau \left( k\right) $ and $\tau
\left( j\right) $ are distinct. Suppose $i\succ _{s}^{J\cup \left\{
k\right\} }j$ and $j\succ _{s}^{J\cup \left\{ i\right\} }k$. By \textbf{PD}, 
$i\succ _{s}^{J}j$ and $j\succ _{s}^{J}k$. Thus, $i\succ _{s}^{J}k$. By 
\textbf{PD}, $i\succ _{s}^{J\cup \left\{ j\right\} }k$.

\subsection*{Proof of Lemma 2.}

Suppose not; that is, for each $i\in J$, there is $j\in J\setminus \left\{
i\right\} $ such that $i\succ _{s}^{J\setminus \left\{ i,j\right\} }j$.
Then, we have an infinite sequence $j_{1},j_{2},\ldots $ such that $j_{n}\in
J$ and $j_{n}\succ _{s}^{J\setminus \left\{ j_{n},j_{n+1}\right\} }j_{n+1}$
for all $n=1,2,\ldots $. Then, since $J$ is finite, $j_{n}=j_{n^{\prime }}$
for some $n,n^{\prime }$ such that $n<n^{\prime }$. Since $j_{n}\succ
_{s}^{J\setminus \left\{ j_{n},j_{n+1}\right\} }j_{n+1}$, $n<n^{\prime }-1$
and thus $j_{n^{\prime }-1}\succ _{s}^{J\setminus \left\{ j_{n^{\prime
}-1},j_{n}\right\} }j_{n}$. By Lemma 1, $j_{n}\succ _{s}^{J\setminus \left\{
j_{n},j_{n+1}\right\} }j_{n+1}$ and $j_{n+1}\succ _{s}^{J\setminus \left\{
j_{n+1},j_{n+2}\right\} }j_{n+2}$ imply $j_{n}\succ _{s}^{J\setminus \left\{
j_{n},j_{n+2}\right\} }j_{n+2}$. Similarly, we have $j_{n}\succ
_{s}^{J\setminus \left\{ j_{n^{\prime }-1},j_{n}\right\} }j_{n^{\prime }-1}$
contradicting $j_{n^{\prime }-1}\succ _{s}^{J\setminus \left\{ j_{n^{\prime
}-1},j_{n}\right\} }j_{n}$. \textbf{Q.E.D.}

\section*{Appendix B}

In this Appendix, we show Theorems 1 and 2.

We introduce the following concepts with regard to choice functions. Fix $%
(P,C)$. First, a matching $\mu $ is $C$-\textbf{stable} if (1) $\mu
R_{i}\emptyset $ for all $i\in I$, (2) $\mu \left( s\right) =C_{s}\left( \mu
\left( s\right) \right) $ for all $s\in S$ and (3) there is no $s$ and $%
i\notin \mu \left( s\right) $ such that $sP_{i}\mu \left( i\right) $ and $%
i\in C_{s}\left( \mu \left( s\right) \cup \left\{ i\right\} \right) $.
Moreover, a matching $\mu $ is said to be \textbf{student-optimal }$C$-%
\textbf{stable} if $\mu $ is stable and not Pareto dominated by any other $C$%
-stable matchings.

We provide two properties on a mechanism with respect to $C$. First, $\phi
^{C}$ is a $C$-\textbf{stable} \textbf{mechanism} if $\phi ^{C}\left(
P\right) $ is $C$-stable for any $P\in \mathcal{P}^{\left\vert I\right\vert
} $. Second, $\phi ^{C}$ is a \textbf{student-optimal }$C$-\textbf{stable
mechanism} if for any $P\in \mathcal{P}^{\left\vert I\right\vert }$, $\phi
^{C}\left( P\right) $ is $C$-stable and $\phi ^{C}\left( P\right) $\ is not
Pareto dominated by any other matchings that are $C$-stable.

We introduce a formal description of the SPDA algorithm with respect to $C$
denoted by $\phi _{DA}^{C}$:

\textbf{Round }$0:$ $J_{s}^{0}=\emptyset $ for all $s$.

\textbf{Round }$d\geq 1:$ Choose one student $i$ who is not tentatively
matched in Round $d-1$ and has not rejected by all acceptable schools for $i$
yet. If there is no such student, then the algorithm terminates. Student $i$
applies to the most preferred $s$ at $P_{i}$ among the schools that have not
rejected $i$ so far. If $i\notin C_{s}\left( J_{s}^{d-1}\cup \left\{
i\right\} \right) $, then $i$ is rejected by $s$, where $J_{s}^{d-1}$ is the
set of students who are tentatively matched to $s$ in the end of Round $d-1$%
. Let $J_{s}^{d}=C_{s}\left( J_{s}^{d-1}\cup \left\{ i\right\} \right) $ and
go to Round $d+1$.

Let $\phi _{DA}^{C}\left( P\right) \left( s\right) =J_{s}^{D}$ for all $s$,
where $D$ is the terminal round of this algorithm.\newline

We provide the following two usual properties on choice functions. First, a
choice function of $s$ denoted by $C_{s}$ is \textbf{substitutable} if for
any $J\subseteq I$, any $i\in I$ and any $j\in I\setminus J$, $i\in
C_{s}\left( J\cup \left\{ j\right\} \right) $ implies $i\in C_{s}\left(
J\right) $. Second, $C_{s}$ is \textbf{size monotonic} if for any $%
J\subseteq J^{\prime }\subseteq I$, $\left\vert C_{s}\left( J\right)
\right\vert \leq \left\vert C_{s}\left( J^{\prime }\right) \right\vert $.

We use the following two well-known results of Roth and Sotomayor (1990) and
Hatfield and Kojima (2009), respectively.

\begin{remark}
(Roth and Sotomayor (1990)) If $C_{s}$ is substitutable for all $s\in S$,
then $\phi _{DA}^{C}$ is a student-optimal $C$-stable mechanism.
\end{remark}

\begin{remark}
(Hatfield and Kojima (2009)) If $C_{s}$ is substitutable and size monotonic
for all $s\in S$, then $\phi _{DA}^{C}$ is a group strategy-proof mechanism.
\end{remark}

Now, we relate stability\textbf{\ }and $C$-stability.

\begin{definition}
A choice function of $s$ denoted by $C_{s}$ is \textbf{consistent with} $%
\succsim _{s}$ if for any $J\subseteq I$, (1) for any $i\in C_{s}\left(
J\right) $ and any $j\in \left( J\setminus C_{s}\left( J\right) \right) \cup
\left\{ \emptyset \right\} $, $i\succsim _{s}^{C_{s}\left( J\right)
\setminus \left\{ i\right\} }j$ and (2) if $\left\vert C_{s}\left( J\right)
\right\vert <q_{s}$ and $\left\vert C_{s}\left( J\right) \right\vert
<\left\vert J\right\vert $, then $\emptyset \succ _{s}^{C_{s}\left( J\right)
}j$ for all $j\in J\setminus C_{s}\left( J\right) $.
\end{definition}

We explain this definition. First, if $i$ is chosen and $j$ is unchosen,
then $j$ does not take priority over $i$ with the chosen students except for 
$i$. Second, if the capacity of $s$ remains and there is an unchosen $j$,
then $j$ is unacceptable for $s$ with the chosen students.

\begin{lemma}
If $C_{s}$ is consistent with $\succsim _{s}$ for all $s\in S$ and $\mu $ is 
$C$-stable, then $\mu $ is also stable.
\end{lemma}

\textbf{Proof. }Suppose that $\mu $ is $C$-stable. We show that $\mu $ is
stable (for $\succsim $). It is straightforward that $\mu $ is non-wasteful
and individually rational. We show that $\mu $ is fair. Suppose not; that
is, $\mu $ is not fair. Then, there are students $i$ and $j$ and school $s$
such that $i\succ _{s}^{\mu \left( s\right) \setminus \left\{ j\right\} }j$, 
$\mu \left( j\right) =s$ and $sP_{i}\mu \left( i\right) $. Since $\mu $ is $%
C $-stable, $j\in C_{s}\left( \mu \left( s\right) \right) =\mu \left(
s\right) $. This implies, by consistency of $C_{s}$, $j\succsim _{s}^{\mu
\left( s\right) \setminus \left\{ j\right\} }$ $\emptyset $. By
transitivity, $i\succsim _{s}^{\mu \left( s\right) \setminus \left\{
j\right\} }\emptyset $.

First, suppose $j\in C_{s}\left( \mu \left( s\right) \cup \left\{ i\right\}
\right) $. We show $i\in C_{s}\left( \mu \left( s\right) \cup \left\{
i\right\} \right) $. Suppose not; that is, $i\notin C_{s}\left( \mu \left(
s\right) \cup \left\{ i\right\} \right) $. Since $i\succ _{s}^{\mu \left(
s\right) \setminus \left\{ j\right\} }j$ and consistency of $C_{s}$, $%
C_{s}\left( \mu \left( s\right) \cup \left\{ i\right\} \right) \neq \mu
\left( s\right) $. Hence $\left\vert C_{s}\left( \mu \left( s\right) \cup
\left\{ i\right\} \right) \right\vert <\left\vert \mu \left( s\right)
\right\vert $, but this contradicts consistency of $C_{s}$ by $i\succsim
_{s}^{\mu \left( s\right) \setminus \left\{ j\right\} }\emptyset $.
Therefore, $j\in C_{s}\left( \mu \left( s\right) \cup \left\{ i\right\}
\right) $ implies $i\in C_{s}\left( \mu \left( s\right) \cup \left\{
i\right\} \right) $.

Second, suppose $j\notin C_{s}\left( \mu \left( s\right) \cup \left\{
i\right\} \right) $. We show $i\in C_{s}\left( \mu \left( s\right) \cup
\left\{ i\right\} \right) $. Suppose not; that is, $i\notin C_{s}\left( \mu
\left( s\right) \cup \left\{ i\right\} \right) $. We have $\left\vert \mu
\left( s\right) \setminus \left\{ j\right\} \right\vert <q_{s}$ and $%
C_{s}\left( \mu \left( s\right) \cup \left\{ i\right\} \right) \subseteq \mu
\left( s\right) \setminus \left\{ j\right\} $. Then, by \textbf{PD}, $%
i\succsim _{s}^{\mu \left( s\right) \setminus \left\{ j\right\} }\emptyset $
implies $i\succsim _{s}^{C_{s}\left( \mu \left( s\right) \cup \left\{
i\right\} \right) }\emptyset $. However, this contradicts consistency of $%
C_{s}$. Therefore, in this case, $i\in C_{s}\left( \mu \left( s\right) \cup
\left\{ i\right\} \right) $ is also satisfied. However, $i\in C_{s}\left(
\mu \left( s\right) \cup \left\{ i\right\} \right) $ contradicts the $C$%
-stability of $\mu $.\textbf{\ Q.E.D.}\newline

By Lemma 3 and Remark 1, we can have a stable matching by using the SPDA
algorithm and a profile of consistent choice functions of all schools, if
they are substitutable. On the other hand, even if $C_{s}$ is consistent
with $\succsim _{s}$ for all $s\in S$ and $\mu $ is stable, $\mu $ may not
be $C$-stable. We introduce an example to show this fact.

Suppose $I=\left\{ 1,2\right\} $, $S=\left\{ s\right\} $, $sP_{1}\emptyset $%
, $sP_{2}\emptyset $, $q_{s}=1$, $1\sim _{s}^{\emptyset }2\succ
_{s}^{\emptyset }\emptyset $, and $2\succ _{s}^{\emptyset }\emptyset $.
Then, $\mu $ such that $\mu \left( 1\right) =s$ and $\mu \left( 2\right)
=\emptyset $ is stable. Next, let $C_{s}$ be such that $C_{s}\left( J\right)
=\left\{ 2\right\} $ if $2\in J,$ and $C_{s}\left( J\right) =\left\{
1\right\} $ if $J=\left\{ 1\right\} $. Then, $C_{s}$ is consistent with $%
\succsim _{s}$. However, $\mu $ is not $C$-stable.

This implies that our mechanism is \textit{not} student-optimally stable in
general, even if $C_{s}$ is consistent with $\succsim _{s}$ and
substitutable for all $s\in S$.

\begin{lemma}
If $C_{s}$ is consistent with $\succsim _{s}$, then it is size monotonic.
\end{lemma}

\textbf{Proof. }Let $J\subseteq J^{\prime }\subseteq I$. If $\left\vert
C_{s}\left( J^{\prime }\right) \right\vert =q_{s}$ or $\left\vert
C_{s}\left( J^{\prime }\right) \right\vert =\left\vert J^{\prime
}\right\vert $, then $\left\vert C_{s}\left( J\right) \right\vert \leq
\left\vert C_{s}\left( J^{\prime }\right) \right\vert $. Therefore, suppose $%
\left\vert C_{s}\left( J^{\prime }\right) \right\vert <q_{s}$ and $%
\left\vert C_{s}\left( J^{\prime }\right) \right\vert <\left\vert J^{\prime
}\right\vert $. Toward a contradiction, we assume $\left\vert C_{s}\left(
J\right) \right\vert >\left\vert C_{s}\left( J^{\prime }\right) \right\vert $%
. Then, there is $t$ such that $\left\vert C_{s}\left( J\right) \cap
I^{t}\right\vert >\left\vert C_{s}\left( J^{\prime }\right) \cap
I^{t}\right\vert $ and thus there is $j\in \left( C_{s}\left( J\right) \cap
I^{t}\right) \setminus C_{s}\left( J^{\prime }\right) $. Since $C_{s}$ is
consistent with $\succsim _{s}$ and $j\notin C_{s}\left( J^{\prime }\right) $%
, $\emptyset \succ _{s}^{C_{s}\left( J^{\prime }\right) }j$. Moreover, since 
$j\in C_{s}\left( J\right) $, $j\succsim _{s}^{C_{s}\left( J\right)
}\emptyset $. However, these contradict \textbf{PD}.\textbf{\ Q.E.D.}\newline

By Lemmata 3 and 4, and Remarks 1 and 2, if $C_{s}$ is consistent with $%
\succsim _{s}$ and substitutable for all $s\in S$, then $\phi _{DA}^{C}$ is
a stable and group strategy-proof mechanism. However, a choice function
which is consistent with $\succsim _{s}$ may not be substitutable. We
introduce an example to show this fact.

Let $I=\{1,2,3\}$ and each of them has a different type from each other. We
assume $q_{s}=1$ and $1\sim _{s}^{\emptyset }2\succsim _{s}^{\emptyset }3$.
Let $C_{s}\left( J\right) =\{2\}$ if $\left\{ 2,3\right\} \subseteq J$ and $%
C_{s}\left( J\right) =\{1\}$ if $\left\vert J\right\vert =2$ and $1\in J$.
Then, $C_{s}$ is consistent with $\succsim _{s}$. However, this is not
substitutable, because $2\notin C_{s}\left( \{1,2\}\right) $ but $2\in
C_{s}\left( \{1,2,3\}\right) $.

Thus, we show that $C_{s}^{\succsim }$, which is defined in Section 4, is
substitutable and consistent with $\succsim _{s}$.

\begin{lemma}
$C_{s}^{\succsim }$ is consistent with $\succsim _{s}$, substitutable and
size monotonic.
\end{lemma}

\textbf{Proof. }To show Lemma 5, the following technical results are
convenient.

\begin{claim}
Let $a_{g}$ and $a_{h}$ be such that $\tau \left( a_{g}\right) =\tau \left(
a_{h}\right) $ with $h>g$. First, if $a_{g}=r_{g^{\prime }}$ where $%
h>g^{\prime }\geq g$, then $g^{\prime }=g$. Second, if $a_{g}=r_{g}$, then $%
a_{h}=r_{h}$. Third, if $a_{g}=r_{g^{\prime }}$ where $g^{\prime }>h>g$,
then $a_{h}=r_{h^{\prime }}$ for some $g^{\prime }>h^{\prime }\geq h$.
Fourth, if $a_{h}$ is not rejected by $s$ in any step, then $a_{g}$ is also
not rejected by $s$ in any step.
\end{claim}

\textbf{Proof of Claim 1.} In this case,%
\begin{equation*}
\tau \left( a_{g}\right) =\tau \left( a_{g+1}\right) =\ldots =\tau \left(
a_{g^{\prime }}\right) =\ldots =\tau \left( a_{h}\right)
\end{equation*}%
and 
\begin{equation*}
a_{g}\succsim _{s}a_{g+1}\succsim _{s}\ldots \succsim _{s}a_{g^{\prime
}}\succsim _{s}\ldots \succsim _{s}a_{h}\text{.}
\end{equation*}%
First, suppose $a_{g}=r_{g^{\prime }}$ where $h>g^{\prime }\geq g$. Suppose
not; that is, $g^{\prime }>g$. Then, $a_{g}$ has the lowest priority among $%
C_{s}^{g^{\prime }-1}\left( J\right) \cup \left\{ a_{g^{\prime }}\right\} $;
that is, 
\begin{equation*}
j\succsim _{s}^{C_{s}^{g^{\prime }-1}\left( J\right) \cup \left\{
a_{g^{\prime }}\right\} \setminus \left\{ j,a_{g}\right\} }a_{g}
\end{equation*}%
for all $j\in C_{s}^{g^{\prime }-1}\left( J\right) \cup \left\{ a_{g^{\prime
}}\right\} $, thus $a_{g}\sim _{s}a_{g^{\prime }}$. By $\tau \left(
a_{g}\right) =\tau \left( a_{g^{\prime }}\right) $ and \textbf{DC},\textbf{\ 
}$j\succsim _{s}^{C_{s}^{g^{\prime }-1}\left( J\right) \setminus \left\{
j\right\} }a_{g^{\prime }}$. Thus, $a_{g^{\prime }}$ has the lowest priority
among $C_{s}^{g^{\prime }-1}\left( J\right) \cup \left\{ a_{g^{\prime
}}\right\} $. Since $g^{\prime }>g$, $a_{g^{\prime }}=r_{g^{\prime }}$ which
is a contradiction. Therefore, if $a_{g}=r_{g^{\prime }}$ where $h>g^{\prime
}\geq g$, then $g^{\prime }=g$.

Second, suppose $a_{g}=r_{g}$. Then, $j\succsim _{s}^{C_{s}^{g-1}\left(
J\right) \setminus \left\{ j\right\} }a_{g}$ for all $j\in C_{s}^{g-1}\left(
J\right) $. By \textbf{DC} and $C_{s}^{g-1}\left( J\right) =C_{s}^{g}\left(
J\right) $, $j\succsim _{s}^{C_{s}^{g}\left( J\right) \setminus \left\{
j\right\} }a_{g+1}$ for all $j\in C_{s}^{g}\left( J\right) .$ Thus, $%
r_{g+1}=a_{g+1}$ and $C_{s}^{g}\left( J\right) =C_{s}^{g+1}\left( J\right) $%
. Similarly, we have $r_{h}=a_{h}$.

Third, suppose $g^{\prime }>h>g$. Suppose not; that is, $a_{h}\in
C_{s}^{g^{\prime }-1}\left( J\right) $. Then, since $a_{g}=r_{g^{\prime }}$
and $g<h$, $a_{h}\succ _{s}^{C_{s}^{g-1}\left( J\right) \cup \left\{
a_{g}\right\} \setminus \left\{ a_{g},a_{h}\right\} }a_{g}$. However, this
contradicts $g<h$, because $g<h$ implies $a_{g}\succsim _{s}a_{h}$.
Therefore, we have $a_{h}=r_{h^{\prime }}$ where $h^{\prime }<g^{\prime }$.

Fourth, suppose that $a_{g}$ is rejected by $s$ in a step $g^{\prime }$.
Then, either $h>g^{\prime }\geq g$ or $g^{\prime }>h>g$ or $g^{\prime }=h$.
However, $g^{\prime }=h$ does not hold, because $a_{g}\succsim _{s}a_{h}$
and $h>g$. Therefore, either $h>g^{\prime }\geq g$ or $g^{\prime }>h>g$.
Then, in either case, $a_{h}$ is rejected in a step. Therefore, by
contraposition, if $a_{h}$ is not rejected by $s$ in any step, then $a_{g}$
is also not rejected by $s$ in any step. \textbf{Q.E.D.}

\begin{claim}
Suppose that a student $r_{g}$ is rejected in Step $g$. If $\left\vert
C_{s}^{h}\left( J\right) \right\vert <q_{s}$ for $h\geq g$, then $\emptyset
\succ _{s}^{C_{s}^{h}\left( J\right) }r_{g}$.
\end{claim}

\textbf{Proof of Claim 2.} Suppose $\left\vert C_{s}^{h}\left( J\right)
\right\vert <q_{s}$. By construction, $C_{s}^{g}\left( J\right) \subseteq
C_{s}^{h}\left( J\right) $ and $\left\vert C_{s}^{g}\left( J\right)
\right\vert \leq \left\vert C_{s}^{h}\left( J\right) \right\vert <q_{s}$.
Therefore, $\emptyset \succ _{s}^{C_{s}^{g}\left( J\right) }r_{g}$.
Moreover, by \textbf{PD }and $C_{s}^{g}\left( J\right) \subseteq
C_{s}^{h}\left( J\right) $, $\emptyset \succ _{s}^{C_{s}^{h}\left( J\right)
}r_{g}$. \textbf{Q.E.D.}

\begin{claim}
Suppose that a student $r_{g}$ is rejected in Step $g$ and $C_{s}^{h}\left(
J\right) \neq \emptyset $ for $h\geq g$. First, for all $i\in
C_{s}^{h}\left( J\right) $, $i\succsim _{s}^{C_{s}^{h}\left( J\right)
\setminus \left\{ i\right\} }r_{g}$. Second, if $i\sim _{s}^{C_{s}^{h}\left(
J\right) \setminus \left\{ i\right\} }r_{g}$ for $i\in C_{s}^{h}\left(
J\right) $, then $\tau \left( i\right) <\tau \left( r_{g}\right) $ or $\tau
\left( i\right) =\tau \left( r_{g}\right) $ and $i<r_{g}$.
\end{claim}

\textbf{Proof of Claim 3}. If $g=h$, these results are trivial. Therefore,
we assume $g<h$.\textbf{\ }

First, we show $i\succsim _{s}^{C_{s}^{h}\left( J\right) \setminus \left\{
i\right\} }r_{g}$ for all $i\in C_{s}^{h}\left( J\right) $ and all $h>g$ by
induction. Suppose $i\succsim _{s}^{C_{s}^{h-1}\left( J\right) \setminus
\left\{ i\right\} }r_{g}$ for all $i\in C_{s}^{h-1}\left( J\right) $. If $%
a_{h}=r_{h}$, then $C_{s}^{h-1}\left( J\right) =C_{s}^{h}\left( J\right) $
and therefore this is trivial. We assume $a_{h}\neq r_{h}$. If $\tau \left(
a_{h}\right) =\tau \left( r_{h}\right) $, then $r_{h}\succsim _{s}a_{h}$ and
thus $a_{h}\,$\ is rejected instead of $r_{h}$. Therefore, $\tau \left(
a_{h}\right) \neq \tau \left( r_{h}\right) $. Since $C_{s}^{h}\left(
J\right) =C_{s}^{h-1}\left( J\right) \cup \left\{ a_{h}\right\} \setminus
\left\{ r_{h}\right\} $, it is sufficient to show 
\begin{equation}
i\succsim _{s}^{C_{s}^{h-1}\left( J\right) \cup \left\{ a_{h}\right\}
\setminus \left\{ r_{h},i\right\} }r_{g}  \label{c1}
\end{equation}%
for all $i\in C_{s}^{h-1}\left( J\right) \cup \left\{ a_{h}\right\}
\setminus \left\{ r_{h}\right\} $. If $\tau \left( i\right) =\tau \left(
r_{g}\right) $, then by \textbf{WO} and $i\succsim _{s}^{C_{s}^{h-1}\left(
J\right) \setminus \left\{ i\right\} }r_{g},$ we have (\ref{c1}).

Suppose $\tau \left( i\right) \neq \tau \left( r_{g}\right) $. By
construction, $i\succsim _{s}^{C_{s}^{h-1}\left( J\right) \cup \left\{
a_{h}\right\} \setminus \left\{ r_{h},i\right\} }r_{h}$ for all $i\in
C_{s}^{h-1}\left( J\right) \cup \left\{ a_{h}\right\} $. Since $r_{h}\in
C_{s}^{h-1}\left( J\right) $, $r_{h}\succsim _{s}^{C_{s}^{h-1}\left(
J\right) \setminus \left\{ r_{h}\right\} }r_{g}$. We show $r_{h}\succsim
_{s}^{C_{s}^{h-1}\left( J\right) \cup \left\{ a_{h}\right\} \setminus
\left\{ r_{h},r_{g}\right\} }r_{g}$. First, suppose $\tau \left( i\right)
=\tau \left( a_{h}\right) $. In this case, by \textbf{PD}, $r_{h}\succsim
_{s}^{C_{s}^{h-1}\left( J\right) \cup \left\{ a_{h}\right\} \setminus
\left\{ r_{h},i\right\} }r_{g}$. Second, suppose $\tau \left( i\right) \neq
\tau \left( a_{h}\right) $. Then, since $\tau \left( i\right) \neq \tau
\left( r_{g}\right) $ and $\tau \left( a_{h}\right) \neq \tau \left(
r_{h}\right) $, we have 
\begin{eqnarray*}
\left\vert C_{s}^{h-1}\left( J\right) \cup \left\{ a_{h}\right\} \setminus
\left\{ r_{h},i\right\} \cap I^{\tau \left( r_{g}\right) }\right\vert &\geq
&\left\vert C_{s}^{h-1}\left( J\right) \setminus \left\{ r_{h}\right\} \cap
I^{\tau \left( r_{g}\right) }\right\vert \text{,} \\
\left\vert C_{s}^{h-1}\left( J\right) \cup \left\{ a_{h}\right\} \setminus
\left\{ r_{h},i\right\} \cap I^{\tau \left( r_{h}\right) }\right\vert &\leq
&\left\vert C_{s}^{h-1}\left( J\right) \setminus \left\{ r_{h}\right\} \cap
I^{\tau \left( r_{h}\right) }\right\vert \text{.}
\end{eqnarray*}%
By \textbf{PD}, we also have $r_{h}\succsim _{s}^{C_{s}^{h-1}\left( J\right)
\cup \left\{ a_{h}\right\} \setminus \left\{ r_{h},i\right\} }r_{g}$. By 
\textbf{NT} and $i\succsim _{s}^{C_{s}^{h-1}\left( J\right) \cup \left\{
a_{h}\right\} \setminus \left\{ r_{h},i\right\} }r_{h}$, we also have (\ref%
{c1}) and thus $i\succsim _{s}^{C_{s}^{h}\left( J\right) \setminus \left\{
i\right\} }r_{g}$ for all $i\in C_{s}^{h}\left( J\right) $.

Second, we show that if $i\sim _{s}^{C_{s}^{h}\left( J\right) \setminus
\left\{ i\right\} }r_{g}$ for $i\in C_{s}^{h}\left( J\right) $, then $\tau
\left( i\right) \leq \tau \left( r_{g}\right) $ by induction. Suppose that
if $i\sim _{s}^{C_{s}^{h-1}\left( J\right) \setminus \left\{ i\right\}
}r_{g} $ for $i\in C_{s}^{h-1}\left( J\right) $, then $\tau \left( i\right)
\leq \tau \left( r_{g}\right) $. If $a_{h}=r_{h}$, then this is trivial.
Therefore, suppose $a_{h}\neq r_{h}$.

We show $i\neq a_{h}$. Since $\tau \left( a_{h}\right) \geq \tau \left(
r_{h}\right) $ and $a_{h}>r_{h}$, $a_{h}\succ _{s}^{C_{s}^{h-1}\left(
J\right) \setminus \left\{ r_{h}\right\} }r_{h}$. Moreover, by the first
result of this claim, $r_{h}\succsim _{s}^{C_{s}^{h-1}\left( J\right)
\setminus \left\{ r_{h}\right\} }r_{g}$. By \textbf{NT}, $a_{h}\succ
_{s}^{C_{s}^{h-1}\left( J\right) \setminus \left\{ r_{h}\right\} }r_{g}$.
Since $C_{s}^{h-1}\left( J\right) \setminus \left\{ r_{h}\right\}
=C_{s}^{h}\left( J\right) \setminus \left\{ a_{h}\right\} $, $a_{h}\succ
_{s}^{C_{s}^{h}\left( J\right) \setminus \left\{ a_{h}\right\} }r_{g}$.
Therefore, if $i\sim _{s}^{C_{s}^{h}\left( J\right) \setminus \left\{
i\right\} }r_{g}$ for $i\in C_{s}^{h}\left( J\right) $, then $i\neq a_{h}$.

Next, we show $\tau \left( i\right) \neq \tau \left( a_{h}\right) $. Suppose
not; that is, $\tau \left( i\right) =\tau \left( a_{h}\right) $. Then, $%
i\succsim _{s}a_{h}$. Moreover, by \textbf{DC }and\textbf{\ }$a_{h}\succ
_{s}^{C_{s}^{h}\left( J\right) \setminus \left\{ a_{h}\right\} }r_{g}$%
\textbf{,} $i\succ _{s}^{C_{s}^{h}\left( J\right) \setminus \left\{
i\right\} }r_{g}$. Therefore, if $i\sim _{s}^{C_{s}^{h}\left( J\right)
\setminus \left\{ i\right\} }r_{g}$ for $i\in C_{s}^{h}\left( J\right) $,
then $\tau \left( i\right) \neq \tau \left( a_{h}\right) $ and $\tau \left(
i\right) <\tau \left( a_{h}\right) $, because of the construction.

Thus, let $i\in C_{s}^{h-1}\left( J\right) $ be such that $i\sim
_{s}^{C_{s}^{h}\left( J\right) \setminus \left\{ i\right\} }r_{g}$ and $\tau
\left( i\right) <\tau \left( a_{h}\right) $. By the first result of this
claim, $i\succsim _{s}^{C_{s}^{h-1}\left( J\right) \setminus \left\{
i\right\} }r_{g}$. We show $\tau \left( i\right) \leq \tau \left(
r_{g}\right) $. First, if $\tau \left( r_{g}\right) \neq \tau \left(
r_{h}\right) $, then%
\begin{eqnarray*}
\left\vert \left( C_{s}^{h-1}\left( J\right) \cup \left\{ a_{h}\right\}
\setminus \left\{ r_{h}\right\} \right) \cap I^{\tau \left( r_{g}\right)
}\right\vert &\geq &\left\vert C_{s}^{h-1}\left( J\right) \cap I^{\tau
\left( r_{g}\right) }\right\vert \\
\left\vert \left( C_{s}^{h-1}\left( J\right) \cup \left\{ a_{h}\right\}
\setminus \left\{ r_{h}\right\} \right) \cap I^{\tau \left( i\right)
}\right\vert &\leq &\left\vert C_{s}^{h-1}\left( J\right) \cap I^{\tau
\left( i\right) }\right\vert ,
\end{eqnarray*}%
\ because $\tau \left( i\right) <\tau \left( a_{h}\right) $, where $%
C_{s}^{h-1}\left( J\right) \cup \left\{ a_{h}\right\} \setminus \left\{
r_{h}\right\} =C_{s}^{h}\left( J\right) $. By \textbf{PD},\textbf{\ }$%
i\succsim _{s}^{C_{s}^{h-1}\left( J\right) \setminus \left\{ i\right\}
}r_{g} $ and $i\sim _{s}^{C_{s}^{h}\left( J\right) \setminus \left\{
i\right\} }r_{g},$ we have $i\sim _{s}^{C_{s}^{h-1}\left( J\right) \setminus
\left\{ i\right\} }r_{g}$. By the construction, we have $\tau \left(
i\right) \leq \tau \left( r_{g}\right) $. Second, suppose $\tau \left(
r_{g}\right) =\tau \left( r_{h}\right) $. Then, by the third result of Claim
4 and $a_{h}\neq r_{h}$, $r_{h}\succsim _{s}r_{g}$. If $\tau \left( i\right)
>\tau \left( r_{g}\right) =\tau \left( r_{h}\right) $, then $i\succ
_{s}^{C_{s}^{h-1}\left( J\right) \cup \left\{ a_{h}\right\} \setminus
\left\{ i,r_{h}\right\} }r_{h}$. In this case, by \textbf{NT,} $i\succ
_{s}^{C_{s}^{h}\left( J\right) \setminus \left\{ i\right\} }r_{g}$, because $%
C_{s}^{h-1}\left( J\right) \cup \left\{ a_{h}\right\} \setminus \left\{
r_{h}\right\} =C_{s}^{h}\left( J\right) $. Hence $\tau \left( i\right) \leq
\tau \left( r_{g}\right) $. Therefore, if $i\sim _{s}^{C_{s}^{h}\left(
J\right) \setminus \left\{ i\right\} }r_{g}$ for $i\in C_{s}^{h}\left(
J\right) $, then $\tau \left( i\right) \leq \tau \left( r_{g}\right) $.
Moreover, by the construction, if $\tau \left( i\right) =\tau \left(
r_{g}\right) $, then $i<r_{g}$. \textbf{Q.E.D.}\newline

\begin{claim}
If $j\in C_{s}^{g}\left( J\right) $, then $j\succsim _{s}^{C_{s}^{h}\left(
J\right) \setminus \left\{ j\right\} }\emptyset $ for all $h\geq g$.
\end{claim}

\textbf{Proof of Claim 4.} If $j\in C_{s}^{g}\left( J\right) $, then $%
j\succsim _{s}^{C_{s}^{g}\left( J\right) \setminus \left\{ j\right\}
}\emptyset $. If $\left\vert C_{s}^{g}\left( J\right) \cap I^{\tau \left(
j\right) }\right\vert \geq \left\vert C_{s}^{h}\left( J\right) \cap I^{\tau
\left( j\right) }\right\vert $, then by \textbf{PD, }$j\succsim
_{s}^{C_{s}^{h}\left( J\right) \setminus \left\{ j\right\} }\emptyset $.
Thus, suppose $\left\vert C_{s}^{g}\left( J\right) \cap I^{\tau \left(
j\right) }\right\vert <\left\vert C_{s}^{h}\left( J\right) \cap I^{\tau
\left( j\right) }\right\vert $. Let $h^{\prime }\in \left( g,h\right] $ be
the smallest integer satisfying $\left\vert C_{s}^{h^{\prime }}\left(
J\right) \cap I^{\tau \left( j\right) }\right\vert =\left\vert
C_{s}^{h}\left( J\right) \cap I^{\tau \left( j\right) }\right\vert $. Then, $%
\tau \left( j\right) =\tau \left( a_{h^{\prime }}\right) $ and $a_{h^{\prime
}}\in C_{s}^{h^{\prime }}\left( J\right) $. Then, $a_{h^{\prime }}\succsim
_{s}^{C_{s}^{h^{\prime }}\left( J\right) \setminus \left\{ a_{h^{\prime
}}\right\} }\emptyset $. Moreover, by the construction, $j\succsim
_{s}a_{h^{\prime }}$. By \textbf{NT,} $j\succsim _{s}^{C_{s}^{h^{\prime
}}\left( J\right) \setminus \left\{ a_{h^{\prime }}\right\} }\emptyset $.
Since $\left\vert C_{s}^{h^{\prime }}\left( J\right) \cap I^{\tau \left(
j\right) }\right\vert =\left\vert C_{s}^{h}\left( J\right) \cap I^{\tau
\left( j\right) }\right\vert $ and \textbf{PD }holds, $j\succsim
_{s}^{C_{s}^{h}\left( J\right) \setminus \left\{ j\right\} }\emptyset $. 
\textbf{Q.E.D.}

\begin{claim}
Let $i\in C_{s}^{\succsim }\left( J\right) $ and $j\in J\setminus
C_{s}^{\succsim }\left( J\right) $. First, $i\succsim _{s}^{C_{s}^{\succsim
}\left( J\right) \setminus \left\{ i\right\} }j$ and $i\succsim
_{s}^{C_{s}^{\succsim }\left( J\right) \setminus \left\{ i\right\}
}\emptyset $. Second, if $\left\vert C_{s}^{\succsim }\left( J\right)
\right\vert <q_{s}$, then $\emptyset \succ _{s}^{C_{s}^{\succsim }\left(
J\right) }j$. Third, if $i\sim _{s}^{C_{s}^{\succsim }\left( J\right)
\setminus \left\{ i\right\} }j$, then $\tau \left( i\right) \leq \tau \left(
j\right) $.
\end{claim}

\textbf{Proof of Claim 5.} First, since $i\in C_{s}^{\succsim }\left(
J\right) $ and $j\in J\setminus C_{s}^{\succsim }\left( J\right) $, $j$ is
rejected at a step. Thus, by the first result of Claim 3, $i\succsim
_{s}^{C_{s}^{\succsim }\left( J\right) \setminus \left\{ i\right\} }j$.
Second, by Claim 4, $i\succsim _{s}^{C_{s}^{\succsim }\left( J\right)
\setminus \left\{ i\right\} }\emptyset $. Third, by Claim 2, if $\left\vert
C_{s}^{\succsim }\left( J\right) \right\vert <q_{s}$, then $\emptyset \succ
_{s}^{C_{s}^{\succsim }\left( J\right) }j$. Finally, by the second result of
Claim 3, if $i\sim _{s}^{C_{s}^{\succsim }\left( J\right) \setminus \left\{
i\right\} }j$, then $\tau \left( i\right) \leq \tau \left( j\right) $. 
\textbf{Q.E.D.}\newline

Now, we show that $C_{s}^{\succsim }$ is consistent with $\succsim _{s}$.
First, suppose $i\in C_{s}^{\succsim }\left( J\right) $ and $j\in J\setminus
C_{s}^{\succsim }\left( J\right) $. By Claim 5, $i\succsim
_{s}^{C_{s}^{\succsim }\left( J\right) \setminus \left\{ i\right\} }j$ and $%
i\succsim _{s}^{C_{s}^{\succsim }\left( J\right) \setminus \left\{ i\right\}
}\emptyset $. Second, suppose $\left\vert C_{s}^{\succsim }\left( J\right)
\right\vert <q_{s}$ and $\left\vert C_{s}^{\succsim }\left( J\right)
\right\vert <\left\vert J\right\vert $. Let $j\in J\setminus C_{s}\left(
J\right) $. By Claim 5, $\emptyset \succ _{s}^{C_{s}^{\succsim }\left(
J\right) }j$.

By Lemma 4, consistency of $C_{s}^{\succsim }$ with $\succsim _{s}$ implies
the size monotonicity of $C_{s}^{\succsim }$.

Next, we show that $C_{s}^{\succsim }$ is substitutable. Let $J\subseteq I$, 
$i\in I$, and $j\in I\setminus J$. We show that $i\in C_{s}^{\succsim
}\left( J\cup \left\{ j\right\} \right) $ implies $i\in C_{s}^{\succsim
}\left( J\right) $. First, we show the following fact.

\begin{claim}
For all $J^{\prime }\subseteq I$ and $j\in I\setminus J^{\prime }$, 
\begin{equation*}
C_{s}^{\succsim }\left( J^{\prime }\cup \left\{ j\right\} \right) \subseteq
C_{s}^{\succsim }\left( C_{s}^{\succsim }\left( J^{\prime }\right) \cup
\left\{ j\right\} \right) .
\end{equation*}
\end{claim}

\textbf{Proof of Claim 6.} For notational simplicity, let $C_{s}^{\succsim
}\left( J^{\prime }\right) \cup \left\{ j\right\} =J^{\ast }$. If $%
C_{s}^{\succsim }\left( J^{\prime }\cup \left\{ j\right\} \right) =\emptyset 
$, then this is obvious. Thus, suppose that there is $i\in C_{s}^{\succsim
}\left( J^{\prime }\cup \left\{ j\right\} \right) $. Toward a contradiction,
suppose $i\notin C_{s}^{\succsim }\left( J^{\ast }\right) $. We consider the
construction process of $C_{s}^{\succsim }$ by letting $J=J^{\prime }\cup
\left\{ j\right\} $. Let $i=a_{h}$. Then, by the fourth result of Claim 1, $%
a_{h^{\prime }}\in C_{s}^{\succsim }\left( J^{\prime }\cup \left\{ j\right\}
\right) $ such that for all $h^{\prime }\leq h$ such that $\tau \left(
a_{h^{\prime }}\right) =\tau \left( i\right) $. Next, we also consider the
construction process of $C_{s}^{\succsim }$ by letting $J=J^{\ast }$. Again,
let $i=a_{h}$. Then, $a_{h}$ is rejected in a process. By contraposition of
the fourth result of Claim 1, $a_{h^{\prime \prime }}\notin C_{s}^{\succsim
}\left( J^{\ast }\right) $ such that for all $h^{\prime \prime }\geq h$ such
that $\tau \left( a_{h^{\prime \prime }}\right) =\tau \left( i\right) $. By
these facts and $J^{\ast }\subseteq J^{\prime }\cup \left\{ j\right\} $, 
\begin{equation}
\left\vert C_{s}^{\succsim }\left( J^{\prime }\cup \left\{ j\right\} \right)
\cap I^{\tau \left( i\right) }\right\vert >\left\vert C_{s}^{\succsim
}\left( J^{\ast }\right) \cap I^{\tau \left( i\right) }\right\vert \text{.}
\label{d1}
\end{equation}%
By Claim 5, $i\succsim _{s}^{C_{s}^{\succsim }\left( J^{\prime }\cup \left\{
j\right\} \right) }\emptyset $. Thus, by \textbf{PD }and (\ref{d1}), $%
i\succsim _{s}^{C_{s}^{\succsim }\left( J^{\ast }\right) }\emptyset $. This
implies $\left\vert C_{s}^{\succsim }\left( J^{\ast }\right) \right\vert
=q_{s}$. Then, 
\begin{equation}
q_{s}=\left\vert C_{s}^{\succsim }\left( J^{\ast }\right) \right\vert \geq
\left\vert C_{s}^{\succsim }\left( J^{\prime }\cup \left\{ j\right\} \right)
\right\vert .  \label{d2}
\end{equation}

Next, by (\ref{d1}) and (\ref{d2}), there is $t\neq \tau \left( i\right) $
such that 
\begin{equation}
\left\vert C_{s}^{\succsim }\left( J^{\prime }\cup \left\{ j\right\} \right)
\cap I^{t}\right\vert <\left\vert C_{s}^{\succsim }\left( J^{\ast }\right)
\cap I^{t}\right\vert \text{.}  \label{d3}
\end{equation}%
Then, there is $i^{\prime }\in C_{s}^{\succsim }\left( J^{\ast }\right) \cap
I^{t}$ but $i^{\prime }\notin C_{s}^{\succsim }\left( J^{\prime }\cup
\left\{ j\right\} \right) $. By the same argument as that above, we have $%
\left\vert C_{s}^{\succsim }\left( J^{\prime }\cup \left\{ j\right\} \right)
\right\vert =q_{s}$.

Then, since $i^{\prime }\in C_{s}^{\succsim }\left( J^{\ast }\right) $, $%
i\notin C_{s}^{\succsim }\left( J^{\ast }\right) $ and Claim 5 is satisfied, 
$i^{\prime }\succsim _{s}^{C_{s}^{\succsim }\left( J^{\ast }\right)
\setminus \left\{ i^{\prime }\right\} }i$. By \textbf{PD,} (\ref{d1}) and (%
\ref{d3}), $i^{\prime }\succsim _{s}^{C_{s}^{\succsim }\left( J^{\prime
}\cup \left\{ j\right\} \right) \setminus \left\{ i\right\} }i$. Since $i\in
C_{s}^{\succsim }\left( J^{\prime }\cup \left\{ j\right\} \right) $ but $%
i^{\prime }\notin C_{s}^{\succsim }\left( J^{\prime }\cup \left\{ j\right\}
\right) $, by Claim 5, $i\sim _{s}^{C_{s}^{\succsim }\left( J^{\prime }\cup
\left\{ j\right\} \right) \setminus \left\{ i\right\} }i^{\prime }$ and
therefore $\tau \left( i^{\prime }\right) >\tau \left( i\right) $. By Claim
5 and $\tau \left( i^{\prime }\right) >\tau \left( i\right) $, $i^{\prime
}\succ _{s}^{C_{s}^{\succsim }\left( J^{\ast }\right) \setminus \left\{
i^{\prime }\right\} }i$. Then, by \textbf{PD,} (\ref{d1}) and (\ref{d3}), $%
i^{\prime }\succ _{s}^{C_{s}^{\succsim }\left( J^{\prime }\cup \left\{
j\right\} \right) \setminus \left\{ i\right\} }i$, which is a contradiction. 
\textbf{Q.E.D. }\newline

Now, we show that for $i\in I$ and $j\in I\setminus J$, $i\in
C_{s}^{\succsim }\left( J\cup \left\{ j\right\} \right) $ implies $i\in
C_{s}^{\succsim }\left( J\right) $. By Claim 6, if $i\in C_{s}^{\succsim
}\left( J^{\prime }\cup \left\{ j\right\} \right) $, then $i\in
C_{s}^{\succsim }\left( C_{s}^{\succsim }\left( J^{\prime }\right) \cup
\left\{ j\right\} \right) $ and thus $i\in C_{s}^{\succsim }\left( J^{\prime
}\right) \cup \left\{ j\right\} $. Since $j\neq i$, $i\in C_{s}^{\succsim
}\left( J^{\prime }\right) $.\textbf{\ Q.E.D. }\newline

Now, we show Theorem 1. By Remark 1 and Lemma 5, $\phi _{DA}^{C^{\succsim
}}\left( P\right) $ is $C^{\succsim }$-stable. Then, by Lemmata 3 and 5, $%
\bar{\phi}\left( P\right) =\phi _{DA}^{C^{\succsim }}\left( P\right) $ is
stable and therefore $\bar{\phi}$ is a stable mechanism. Next, Remark 2 and
Lemma 5 imply that $\phi _{DA}^{C^{\succsim }}$ is group strategy-proof.
Therefore, $\bar{\phi}$ is group strategy-proof.\newline

Next, to show Theorem 2, we introduce the following result.

\begin{lemma}
Suppose that $\succsim _{s}^{J}$ is a linear order for all $J\in \mathcal{J}%
_{s}$ and all $s$, and $C_{s}$ is consistent with $\succsim _{s}$ for all $s$%
. Then, $\mu $ is $C$-stable if and only if $\mu $ is stable.
\end{lemma}

\textbf{Proof.} Suppose that $C_{s}$ is consistent with $\succsim _{s}$ for
all $s\in S$. By Lemma 3, it is sufficient to show that if $\mu $ is stable,
then $\mu $ is also $C$-stable. Suppose that $\mu $ is stable. Then, $\mu
\left( i\right) R_{i}\emptyset $ for all $i\in I$. First, we show $\mu
\left( s\right) =C_{s}\left( \mu \left( s\right) \right) $ for all $s\in S$.
Suppose not; that is, $C_{s}\left( \mu \left( s\right) \right) \subsetneq
\mu \left( s\right) $. Then, we can let $i\in \mu \left( s\right) \setminus
C_{s}\left( \mu \left( s\right) \right) $. Since $C_{s}$ is consistent with $%
\succsim _{s}$ and $\succ _{s}^{\mu \left( s\right) \setminus \left\{
i\right\} }$ is a linear order, $i\succ _{s}^{\mu \left( s\right) \setminus
\left\{ i\right\} }\emptyset $. By $C_{s}\left( \mu \left( s\right) \right)
\subsetneq \mu \left( s\right) $ and \textbf{PD}, $i\succ _{s}^{C_{s}\left(
\mu \left( s\right) \right) }\emptyset $. Since $\left\vert C_{s}\left( \mu
\left( s\right) \right) \right\vert <\left\vert \mu \left( s\right)
\right\vert $, $\left\vert C_{s}\left( \mu \left( s\right) \right)
\right\vert =q_{s}$. However, this contradicts $q_{s}\geq \left\vert \mu
\left( s\right) \right\vert $. Therefore, $\mu \left( s\right) =C_{s}\left(
\mu \left( s\right) \right) $ for all $s\in S$.

Second, we show that there is no pair $(s,i)$ such that $sP_{i}\mu \left(
i\right) $ and $i\in C_{s}\left( \mu \left( s\right) \cup \left\{ i\right\}
\right) $. Suppose not; that is, there is $(s,i)$ such that $sP_{i}\mu
\left( i\right) $ and $i\in C_{s}\left( \mu \left( s\right) \cup \left\{
i\right\} \right) $. First, suppose $\emptyset \succ _{s}^{\mu \left(
s\right) }i$. By $i\in C_{s}\left( \mu \left( s\right) \cup \left\{
i\right\} \right) $, $i\succsim _{s}^{C_{s}\left( \mu \left( s\right) \cup
\left\{ i\right\} \right) \setminus \left\{ i\right\} }\emptyset $. By 
\textbf{PD}, these imply that there is $i^{\prime }$ $\in \mu \left(
s\right) \setminus C_{s}\left( \mu \left( s\right) \cup \left\{ i\right\}
\right) $ such that $\tau \left( i^{\prime }\right) =\tau \left( i\right) $.
Then, by \textbf{WO }and $i\in C_{s}\left( \mu \left( s\right) \cup \left\{
i\right\} \right) $ and $i^{\prime }$ $\notin C_{s}\left( \mu \left(
s\right) \cup \left\{ i\right\} \right) $, we have $i\succsim _{s}i^{\prime
} $. Since $\succsim _{s}$ is a linear order, $i\succ _{s}i^{\prime }$.
Since $i^{\prime }\in \mu \left( s\right) $, $i$ has justified envy toward $%
i^{\prime }$ at $\mu $ contradicting the stability of $\mu $.

Therefore, $i\succ _{s}^{\mu \left( s\right) }\emptyset $. Since $\mu $ is
stable and $i\succ _{s}^{\mu \left( s\right) }\emptyset $, $\left\vert \mu
\left( s\right) \right\vert =q_{s}$ and thus there is $j\in \mu \left(
s\right) \setminus C_{s}\left( \mu \left( s\right) \cup \left\{ i\right\}
\right) $. Then, $i\succ _{s}^{C_{s}\left( \mu \left( s\right) \cup \left\{
i\right\} \right) \setminus \left\{ i\right\} }j$ and $j\succ _{s}^{\mu
\left( s\right) \setminus \left\{ j\right\} }i$. Since $C_{s}\left( \mu
\left( s\right) \cup \left\{ i\right\} \right) \setminus \left\{ i\right\}
\subseteq \mu \left( s\right) $ and \textbf{PD} is satisfied, there is $%
i^{\prime \prime }$ such that $i^{\prime \prime }$ $\in \mu \left( s\right)
\setminus C_{s}\left( \mu \left( s\right) \cup \left\{ i\right\} \right) $
such that $\tau \left( i^{\prime \prime }\right) =\tau \left( i\right) $. By 
\textbf{WO }and $\succsim _{s}^{J}$ is a linear order for all $J\in \mathcal{%
J}_{s}$, $i\in C_{s}\left( \mu \left( s\right) \cup \left\{ i\right\}
\right) $ and $i^{\prime \prime }\notin C_{s}\left( \mu \left( s\right) \cup
\left\{ i\right\} \right) $, we have $i\succsim _{s}i^{\prime \prime }$.
Since $\succsim _{s}^{J}$ is a linear order for all $J\in \mathcal{J}_{s}$, $%
i\succ _{s}i^{\prime \prime }$. Since $i^{\prime \prime }\in \mu \left(
s\right) $, $i$ has justified envy toward $i^{\prime \prime }$ at $\mu $,
contradicting the stability of $\mu $. \textbf{Q.E.D.}\newline

Now, we show Theorem 2; that is, we show the student-optimal stability of $%
\bar{\phi}$ in this case. Suppose not; that is, there is a pair $\left(
P,\succsim \right) $ such that there is a stable matching $\mu $ that Pareto
dominates $\bar{\phi}\left( P\right) $. By Lemma 6, since $C_{s}^{\succsim }$
is consistent with $\succsim $, then $\mu $ is $C^{\succsim }$-stable.
Therefore, $\mu $ Pareto dominates $\phi _{DA}^{C^{\succsim }}\left(
P\right) $ meaning that $\phi _{DA}^{C^{\succsim }}$ is not student-optimal $%
C$-stable mechanism contradicting Remark 2 and Lemma 5.

\section*{Appendix C}

In this Appendix, we show that our model is a generalization of several
models in previous studies. First, we consider two choice functions
introduced by Echenique and Yenmez (2015). We consider the adjusted scoring
rules, where $\sigma _{1},\sigma _{2},\ldots ,\sigma _{\left\vert
I\right\vert }$ are distinct. We show that for each of the two choice
functions introduced by Echenique and Yenmez (2015), there is $\left( 
\underline{\sigma },\alpha \right) \in \mathbb{R}\times A$ such that if $%
\succsim _{s}$ is an adjusted scoring rule with respect to $\left( \sigma ,%
\underline{\sigma },\alpha \right) \in \left[ 0,1\right] ^{I}\times \mathbb{R%
}\times A$, then the choice function is consistent with $\succsim _{s}$.

\begin{definition}
A choice function $C_{s}$ is generated by reserves for priority determined
by $\sigma $ if there exists a vector $r=\left( r_{t}\right) _{t\in T}\in 
\mathbb{Z}_{++}^{_{\left\vert T\right\vert }}$ with $\sum_{t\in T}r_{t}\leq
q_{s}$ such that for any $J\subseteq I,$ (i) $\left\vert C_{s}\left(
J\right) \cap I^{t}\right\vert \geq \min \left\{ r_{t},\left\vert J\cap
I^{t}\right\vert \right\} ,$ (ii) if $i\in C_{s}\left( J\right) $, $j\in
J\setminus C_{s}\left( J\right) $, and $\sigma _{j}>\sigma _{i}$, then $\tau
\left( i\right) \neq \tau \left( j\right) $ and $\left\vert C_{s}\left(
J\right) \cap I^{\tau \left( i\right) }\right\vert \leq r_{\tau \left(
i\right) },$ (iii) if $J\setminus C_{s}\left( J\right) \neq \emptyset $,
then $\left\vert C_{s}\left( J\right) \right\vert =q_{s}$.
\end{definition}

\begin{definition}
A choice function $C_{s}$ is generated by quotas for priority determined by $%
\sigma $ if there exists a vector a vector $r=\left( r_{t}\right) _{t\in
T}\in \mathbb{Z}_{++}^{_{\left\vert T\right\vert }}$ such that for any $%
J\subseteq I,$ (i) $\left\vert C_{s}\left( J\right) \cap I^{t}\right\vert
\leq r_{t}$ (ii) if $i\in C_{s}\left( J\right) $, $j\in J\setminus
C_{s}\left( J\right) $, and $\sigma _{j}>\sigma _{i}$, then $\tau \left(
i\right) \neq \tau \left( j\right) $ and $\left\vert C_{s}\left( J\right)
\cap I^{\tau \left( j\right) }\right\vert =$\ $r_{\tau \left( j\right) },$
(iii) if $i\in J\setminus C_{s}\left( J\right) $, then either $\left\vert
C_{s}\left( J\right) \right\vert =q_{s}$ or $\left\vert C_{s}\left( J\right)
\cap I^{\tau \left( i\right) }\right\vert =$\ $r_{\tau \left( i\right) }$.
\end{definition}

Let 
\begin{eqnarray}
\alpha _{\tau \left( i\right) }\left( x\right) &=&2\text{ if }r_{\tau \left(
i\right) }>x\text{,}  \notag \\
&=&0\text{ if }x\geq r_{\tau \left( i\right) },  \label{x}
\end{eqnarray}%
for all $i\in I$ and all $x\in \mathbb{Z}_{++}$. Then, we have the following
results.

\begin{proposition}
Suppose that $\sigma _{1},\sigma _{2},\ldots ,\sigma _{\left\vert
I\right\vert }$ are distinct. Let $C_{s}$ be a choice function generated by
reserves for priority determined by $\sigma $. If $\succsim _{s}$ is an
adjusted scoring rule with respect to $\left( \sigma ,\underline{\sigma }%
,\alpha \right) \in \left[ 0,1\right] ^{I}\times \mathbb{R}\times A$
satisfying (\ref{x}) and $\underline{\sigma }<0$, then $\succsim _{s}^{J}$
is a linear order for all $J\in \mathcal{J}_{s}$ and $C_{s}$ is consistent
with $\succsim _{s}$.
\end{proposition}

\textbf{Proof.} First, since $\sigma _{1},\sigma _{2},\ldots ,\sigma
_{\left\vert I\right\vert }$ are distinct, $\succsim _{s}^{J}$ is a linear
order for any $J\in \mathcal{J}_{s}$.

We show that for any $i\in C_{s}\left( J\right) $ and any $j\in \left(
J\setminus C_{s}\left( J\right) \right) \cup \left\{ \emptyset \right\} $, $%
i\succsim _{s}^{C_{s}\left( J\right) \setminus \left\{ i\right\} }j$.
Suppose not; that is, for some $i\in C_{s}\left( J\right) $ and some $j\in
\left( J\setminus C_{s}\left( J\right) \right) \cup \left\{ \emptyset
\right\} $, $j\succ _{s}^{C_{s}\left( J\right) \setminus \left\{ i\right\}
}i $. Since (\ref{x}) is satisfied, $f_{i}\left( J\right) \geq 0$ for any $%
J\subseteq I$. Since $\underline{\sigma }<0$, $i\succsim _{s}^{J}\emptyset $
for all $J$ and all $i\in I\setminus J$. Therefore, $j\neq \emptyset $.
Moreover, $j\succ _{s}^{C_{s}\left( J\right) \setminus \left\{ i\right\} }i$
implies 
\begin{equation}
f_{j}\left( C_{s}\left( J\right) \setminus \left\{ i\right\} \right)
>f_{i}\left( C_{s}\left( J\right) \setminus \left\{ i\right\} \right) \text{.%
}  \label{B1}
\end{equation}

First, suppose $\sigma _{i}>\sigma _{j}$. By (\ref{x}), $\left\vert
C_{s}\left( J\setminus \left\{ i\right\} \right) \cap I^{\tau \left(
j\right) }\right\vert <r_{\tau \left( j\right) }$. Moreover, $j\in
J\setminus C_{s}\left( J\right) $ implies $\left\vert C_{s}\left( J\setminus
\left\{ i\right\} \right) \cap I^{\tau \left( j\right) }\right\vert
<\left\vert J\cap I^{\tau \left( j\right) }\right\vert $. Then, these
contradict (i). Therefore, $\sigma _{j}>\sigma _{i}$. By (ii), $\tau \left(
i\right) \neq \tau \left( j\right) $ and $\left\vert C_{s}\left( J\right)
\cap I^{\tau \left( i\right) }\right\vert \leq r_{\tau \left( i\right) }$.
Then, (\ref{x}) and (\ref{B1}) imply $\left\vert C_{s}\left( J\setminus
\left\{ i\right\} \right) \cap I^{\tau \left( j\right) }\right\vert <r_{\tau
\left( j\right) }$ contradicting (i), because $j\in J\setminus C_{s}\left(
J\right) $. Therefore, for any $i\in C_{s}\left( J\right) $ and any $j\in
\left( J\setminus C_{s}\left( J\right) \right) \cup \left\{ \emptyset
\right\} $, $i\succsim _{s}^{C_{s}\left( J\right) \setminus \left\{
i\right\} }j$.

Second, we show that if $\left\vert C_{s}\left( J\right) \right\vert <q_{s}$
and $\left\vert C_{s}\left( J\right) \right\vert <\left\vert J\right\vert $,
then $\emptyset \succ _{s}^{C_{s}\left( J\right) }j$ for all $j\in
J\setminus C_{s}\left( J\right) $. If $\left\vert C_{s}\left( J\right)
\right\vert <\left\vert J\right\vert $, then $J\setminus C_{s}\left(
J\right) \neq \emptyset $. In this case, by (iii), $\left\vert C_{s}\left(
J\right) \right\vert =q_{s}$. Therefore, $\left\vert C_{s}\left( J\right)
\right\vert <\left\vert J\right\vert $ is incompatible with $\left\vert
C_{s}\left( J\right) \right\vert <q_{s}$. Therefore, if $\left\vert
C_{s}\left( J\right) \right\vert <q_{s}$ and $\left\vert C_{s}\left(
J\right) \right\vert <\left\vert J\right\vert $, then $\emptyset \succ
_{s}^{C_{s}\left( J\right) }j$ for all $j\in J\setminus C_{s}\left( J\right) 
$. \textbf{Q.E.D.}\newline

By Proposition 2 and Lemma 6, when $C_{s}$ is generated by reserves for
priority determined by $\sigma $ for all $s\in S$, a matching is $C$-stable
if and only if it is stable (under $\succsim $).

\begin{proposition}
Suppose that $\sigma _{1},\sigma _{2},\ldots ,\sigma _{\left\vert
I\right\vert }$ are distinct. Let $C_{s}$ be generated by quotas for
priority determined by $\sigma $. If $\succsim _{s}$ is an adjusted scoring
rule with respect to $\left( \sigma ,\underline{\sigma },\alpha \right) \in %
\left[ 0,1\right] ^{I}\times \mathbb{R}\times A$ satisfying (\ref{x}) and $%
\underline{\sigma }=2$, then $\succsim _{s}^{J}$ is a linear order for all $%
J\in \mathcal{J}_{s}$ and $C_{s}$ is consistent with $\succsim _{s}$.
\end{proposition}

\textbf{Proof. }First, since $\sigma _{1},\sigma _{2},\ldots ,\sigma
_{\left\vert I\right\vert }$ are distinct, $\succsim _{s}^{J}$ is a linear
order for any $J\in \mathcal{J}_{s}$.

Next, we show for any $i\in C_{s}\left( J\right) $ and any $j\in \left(
J\setminus C_{s}\left( J\right) \right) \cup \left\{ \emptyset \right\} $, $%
i\succsim _{s}^{C_{s}\left( J\right) \setminus \left\{ i\right\} }j$.
Suppose not; that is, for some $i\in C_{s}\left( J\right) $ and some $j\in
\left( J\setminus C_{s}\left( J\right) \right) \cup \left\{ \emptyset
\right\} $, $j\succ _{s}^{C_{s}\left( J\right) \setminus \left\{ i\right\}
}i $. By (i), $\left\vert C_{s}\left( J\right) \cap I^{\tau \left( i\right)
}\right\vert \leq r_{\tau \left( i\right) }$ implies $\left\vert C_{s}\left(
J\right) \setminus \left\{ i\right\} \cap I^{\tau \left( i\right)
}\right\vert <r_{\tau \left( i\right) }$. By (\ref{x}) and $\underline{%
\sigma }=2$, $i\succsim _{s}^{C_{s}\left( J\right) \setminus \left\{
i\right\} }\emptyset $. Hence $j\neq \emptyset $. Now, suppose $\sigma
_{i}>\sigma _{j}$. Then, by (\ref{x}), $\left\vert C_{s}\left( J\right)
\setminus \left\{ i\right\} \cap I^{\tau \left( i\right) }\right\vert \geq
r_{\tau \left( i\right) }$ and thus, $\left\vert C_{s}\left( J\right) \cap
I^{\tau \left( i\right) }\right\vert >r_{\tau \left( i\right) }$
contradicting (i). Hence $\sigma _{j}>\sigma _{i}$. By (ii), $\tau \left(
i\right) \neq \tau \left( j\right) $ and $\left\vert C_{s}\left( J\setminus
\left\{ i\right\} \right) \cap I^{\tau \left( j\right) }\right\vert =$\ $%
r_{\tau \left( j\right) }$. Moreover, by (i), $\left\vert C_{s}\left(
J\setminus \left\{ i\right\} \right) \cap I^{\tau \left( i\right)
}\right\vert <r_{\tau \left( i\right) }$. By (\ref{x}), $f_{i}\left(
C_{s}\left( J\right) \setminus \left\{ i\right\} \right) >f_{j}\left(
C_{s}\left( J\right) \setminus \left\{ i\right\} \right) $ and thus $i\succ
_{s}^{C_{s}\left( J\right) \setminus \left\{ i\right\} }j$ which is a
contradiction.

Second, we show that if $\left\vert C_{s}\left( J\right) \right\vert <q_{s}$
and $\left\vert C_{s}\left( J\right) \right\vert <\left\vert J\right\vert $,
then $\emptyset \succ _{s}^{C_{s}\left( J\right) }j$ for all $j\in
J\setminus C_{s}\left( J\right) $. Suppose $\left\vert C_{s}\left( J\right)
\right\vert <q_{s}$ and $\left\vert C_{s}\left( J\right) \right\vert
<\left\vert J\right\vert $. Then, by (iii), for all $j\in J\setminus
C_{s}\left( J\right) $, $\left\vert C_{s}\left( J\right) \cap I^{\tau \left(
j\right) }\right\vert =$\ $r_{\tau \left( j\right) }$. By (\ref{x}), $%
f_{j}\left( C_{s}\left( J\right) \right) <2=\underline{\sigma }$ and thus $%
\emptyset \succ _{s}^{C_{s}\left( J\right) }j$. \textbf{Q.E.D. }\newline

By Proposition 3 and Lemma 6, when $C_{s}$ is generated by quotas for
priority determined by $\sigma $ for all $s\in S$, a matching is $C$-stable
if and only if it is stable (under $\succsim $).

Second, we show that our model is a generalization of the controlled school
choice model with soft bounds introduced by Ehlers et al. (2014). Let $\rho
=\left( \rho _{t}\right) _{t\in T}\in \mathbb{Z}_{+}^{_{\left\vert
T\right\vert }}$, satisfying $\rho _{t}\geq r_{t}$. The following concept of
fairness is introduced by Ehlers et al. (2014).

\begin{definition}
A matching $\mu $ is \textbf{fair under soft bounds} if for any student $i$
and any school $s$ such that $sP_{i}\mu \left( i\right) $ with $\tau \left(
i\right) =t$, we have $\sigma _{j}\geq \sigma _{i}$ for all $j\in \mu \left(
s\right) \cap I^{t}$, and either (i) $\left\vert \mu \left( s\right) \cap
I^{t}\right\vert \geq \rho _{t}$ and $\sigma _{j}\geq \sigma _{i}$ for all $%
j\in \mu \left( s\right) $ such that $\left\vert \mu \left( s\right) \cap
I^{\tau \left( j\right) }\right\vert >\rho _{\tau \left( j\right) }$, or
(ii) $\left\vert \mu \left( s\right) \cap I^{t}\right\vert \in \left[
r_{t},\rho _{t}\right) $, and (iia) $\left\vert \mu \left( s\right) \cap
I^{t^{\prime }}\right\vert \leq \rho _{t^{\prime }}$ for all $t^{\prime }$ $%
\in T\setminus \{t\}$, and (iib) $\sigma _{j}\geq \sigma _{i}$ for all $j\in
\mu \left( s\right) $ such that $\left\vert \mu \left( s\right) \cap I^{\tau
\left( j\right) }\right\vert \in \left( r_{\tau \left( j\right) },\rho
_{\tau \left( j\right) }\right] $ or (iii) $\left\vert \mu \left( s\right)
\cap I^{t}\right\vert <r_{t}$, $\left\vert \mu \left( s\right) \cap I^{\tau
\left( j\right) }\right\vert \leq r_{\tau \left( j\right) }$ and $\sigma
_{j}\geq \sigma _{i}$ for all $j\in \mu \left( s\right) $.\footnote{%
Ehlers et al. (2014) require that any student $i$ and any school $s$ such
that $sP_{i}\mu \left( i\right) $ with $\tau \left( i\right) =t$, $%
\left\vert \mu \left( s\right) \cap I^{t}\right\vert \geq r_{t}$. However,
if for all $j\in \mu \left( s\right) $, $\left\vert \mu \left( s\right) \cap
I^{\tau \left( j\right) }\right\vert \leq r_{\tau \left( j\right) }$ and $%
\sigma _{j}\geq \sigma _{i}$, $i$ has no justified envy at $s$ with $\mu $.
The result of Ehlers et al. (2014, Theorem 4) is not changed with this
definition, because $\left\vert \mu \left( s\right) \cap I^{t}\right\vert
\geq r_{t}$ is satisfied if $\mu $ is nonwasteful under soft bounds and fair
under soft bounds in either definition.}
\end{definition}

We show that there exist some $\left( \underline{\sigma },\alpha \right) $
such that, if $\succsim _{s}$ is an adjusted scoring rule with respect to $%
\left( \sigma ,\underline{\sigma },\alpha \right) $ for all $s\in S$, then a
matching $\mu $ is fair if and only if it is fair under soft bounds. Let 
\begin{eqnarray}
\alpha _{\tau \left( i\right) }\left( x\right) &=&4\text{ if }x<r_{\tau
\left( i\right) },  \notag \\
&=&2\text{ if }x\in \left[ r_{\tau \left( i\right) },\rho _{\tau \left(
i\right) }\right) ,  \notag \\
&=&0\text{ if }x\geq \rho _{\tau \left( i\right) }\text{ }  \label{y}
\end{eqnarray}%
for all $x\in \mathbb{Z}_{++}$, $i\in I$ and $\underline{\sigma }<0$.

\begin{proposition}
Suppose that for all $s\in S$, $\succsim _{s}$ is an adjusted scoring rule
with respect to $\left( \sigma ,\underline{\sigma },\alpha \right) \in \left[
0,1\right] ^{I}\times \mathbb{R}\times A$ satisfying (\ref{y}), $\sum_{t\in
T}r_{t}\leq q_{s}$ and $\underline{\sigma }<0$. Then, a matching $\mu $ is
fair (for $\succsim _{s}$) if and only if it is fair under soft bounds.
\end{proposition}

\textbf{Proof. }First, we show that if $i\in I\setminus \mu \left( s\right) $
has justified envy toward $j\in \mu \left( s\right) $ at $\mu $; that is, if 
$sP_{i}\mu \left( i\right) $ and $i\succ _{s}^{\mu \left( s\right) \setminus
\left\{ j\right\} }j$, then $\mu $ is not fair under soft bounds. First,
suppose $\tau \left( i\right) =\tau \left( j\right) $. Then, since $i\succ
_{s}^{\mu \left( s\right) \setminus \left\{ j\right\} }j$\ and 
\begin{equation*}
\alpha _{\tau \left( i\right) }\left( \left\vert \left( \mu \left( s\right)
\setminus \left\{ j\right\} \right) \cap I^{\tau \left( i\right)
}\right\vert \right) =\alpha _{\tau \left( j\right) }\left( \left\vert
\left( \mu \left( s\right) \setminus \left\{ j\right\} \right) \cap I^{\tau
\left( j\right) }\right\vert \right) ,
\end{equation*}%
$\sigma _{i}>\sigma _{j}$. Therefore, $\mu $ is not fair under soft bounds.
Hereafter, we assume $\tau \left( i\right) \neq \tau \left( j\right) $.
Then, $\alpha _{\tau \left( i\right) }\left( \left\vert \left( \mu \left(
s\right) \setminus \left\{ j\right\} \right) \cap I^{\tau \left( i\right)
}\right\vert \right) =\alpha _{\tau \left( i\right) }\left( \left\vert \mu
\left( s\right) \cap I^{\tau \left( i\right) }\right\vert \right) $.

First, we assume $\left\vert \mu \left( s\right) \cap I^{\tau \left(
i\right) }\right\vert \geq \rho _{\tau \left( i\right) }$. Then, by (\ref{y}%
), $f_{i}\left( \left\vert \left( \mu \left( s\right) \setminus \left\{
j\right\} \right) \cap I^{\tau \left( i\right) }\right\vert \right) =\sigma
_{i}$. Then, $i\succ _{s}^{\mu \left( s\right) \setminus \left\{ j\right\}
}j $ implies $f_{j}\left( \left\vert \left( \mu \left( s\right) \setminus
\left\{ j\right\} \right) \cap I^{\tau \left( j\right) }\right\vert \right)
<\sigma _{i}$ and therefore, $\left\vert \left( \mu \left( s\right)
\setminus \left\{ j\right\} \right) \cap I^{\tau \left( j\right)
}\right\vert \geq \rho _{\tau \left( j\right) }$ and $\sigma _{j}<\sigma
_{i} $. This implies that $\mu $ is not fair under soft bounds. Second, we
assume $\left\vert \left( \mu \left( s\right) \setminus \left\{ j\right\}
\right) \cap I^{\tau \left( i\right) }\right\vert \in \left[ r_{\tau \left(
i\right) },\rho _{\tau \left( i\right) }\right) $. Since 
\begin{equation*}
f_{i}\left( \left\vert \left( \mu \left( s\right) \setminus \left\{
j\right\} \right) \cap I^{\tau \left( i\right) }\right\vert \right)
=2+\sigma _{i}>f_{j}\left( \left\vert \left( \mu \left( s\right) \setminus
\left\{ j\right\} \right) \cap I^{\tau \left( j\right) }\right\vert \right) ,
\end{equation*}%
$\left\vert \left( \mu \left( s\right) \setminus \left\{ j\right\} \right)
\cap I^{\tau \left( j\right) }\right\vert \geq r_{\tau \left( j\right) }$.
If $\left\vert \left( \mu \left( s\right) \setminus \left\{ j\right\}
\right) \cap I^{\tau \left( j\right) }\right\vert \geq \rho _{\tau \left(
j\right) }$; that is, if $\left\vert \mu \left( s\right) \cap I^{\tau \left(
j\right) }\right\vert >\rho _{\tau \left( j\right) }$, then (iia) is not
satisfied and thus $\mu $ is not fair under soft bounds. Hence we assume $%
\left\vert \mu \left( s\right) \setminus \left\{ j\right\} \cap I^{\tau
\left( j\right) }\right\vert \in \left[ r_{\tau \left( j\right) },\rho
_{\tau \left( j\right) }\right) $. Then $\sigma _{j}<\sigma _{i}$ and thus
(iib) is not satisfied. Thus, $\mu $ is not fair under soft bounds. Third,
suppose $\left\vert \mu \left( s\right) \cap I^{\tau \left( i\right)
}\right\vert <r_{\tau \left( i\right) }$. Then, $i\succ _{s}^{\mu \left(
s\right) \setminus \left\{ j\right\} }j$ implies $\left\vert \left( \mu
\left( s\right) \setminus \left\{ j\right\} \right) \cap I^{\tau \left(
i\right) }\right\vert \geq r_{\tau \left( i\right) }$ or $\sigma _{j}<\sigma
_{i}$. Hence $\mu $ is not fair under soft bounds.

Second, suppose that $\mu $ is not fair under soft bounds. Then, there are $%
i $ and $s$ such that $sP_{i}\mu \left( i\right) $ with $\tau \left(
i\right) =t$, $\sigma _{j}<\sigma _{i}$ for some $j\in \mu \left( s\right)
\cap I^{t}$ or any of (i), (ii) and (iii) is not satisfied. We show that $i$
has justified envy toward some $j\in \mu \left( s\right) $ at $\mu $. First,
if $\sigma _{j}<\sigma _{i}$ for some $j\in \mu \left( s\right) \cap I^{t}$,
then $i\succ _{s}^{\mu \left( s\right) \setminus \left\{ j\right\} }j$.
Thus, $i$ has justified envy toward $j$ at $\mu $. We assume $\sigma
_{j}\geq \sigma _{i}$ for any $j\in \mu \left( s\right) \cap I^{t}$. Then,
any of (i), (ii) and (iii) is not satisfied. First, suppose $\left\vert \mu
\left( s\right) \cap I^{t}\right\vert \geq \rho _{t}$. Since (i) is not
satisfied, $\sigma _{j}<\sigma _{i}$ for some $j\in \mu \left( s\right) $
such that $\left\vert \mu \left( s\right) \cap I^{\tau \left( j\right)
}\right\vert >\rho _{\tau \left( j\right) }$. Then, by (\ref{y}), $i\succ
_{s}^{\mu \left( s\right) \setminus \left\{ j\right\} }j$. Second, suppose $%
\left\vert \mu \left( s\right) \cap I^{t}\right\vert \in \left[ r_{t},\rho
_{t}\right) $. If (iia) is not satisfied; that is, if $\left\vert \mu \left(
s\right) \cap I^{t^{\prime }}\right\vert >\rho _{t^{\prime }}$ for some $%
t^{\prime }$ $\in T\setminus \{t\}$, then there is $j\in \mu \left( s\right)
\cap I^{t^{\prime }}$ such that $j\succ _{s}^{\mu \left( s\right) \setminus
\left\{ j\right\} }i$. If (iib) is not satisfied; that is, if $\sigma
_{j}<\sigma _{i}$ for some $j\in \mu \left( s\right) $ such that $\left\vert
\mu \left( s\right) \cap I^{\tau \left( j\right) }\right\vert \in \left(
r_{\tau \left( j\right) },\rho _{\tau \left( j\right) }\right] $, then $%
j\succ _{s}^{\mu \left( s\right) \setminus \left\{ j\right\} }i$. Third,
suppose $\left\vert \mu \left( s\right) \cap I^{t}\right\vert <r_{t}$. Since
(iii) is not satisfied, either $\left\vert \mu \left( s\right) \cap I^{\tau
\left( j\right) }\right\vert >r_{\tau \left( j\right) }$ or $\sigma
_{j}<\sigma _{i}$ for some $j\in \mu \left( s\right) $. In either case, $%
j\succ _{s}^{\mu \left( s\right) \setminus \left\{ j\right\} }i$. \textbf{%
Q.E.D.}\newline

Finally, we briefly consider the reserves-and-quotas rule introduced by
Imamura (2023). He mentions that the rule can be regarded as that with hard
upper and soft lower type-specific bounds in the terminology of Ehlers et
al. (2014). Thus, we let $\succsim _{s}$ be an adjusted scoring rule with
respect to $\left( \sigma ,\underline{\sigma },\alpha \right) $ satisfying (%
\ref{y}) and $\underline{\sigma }=2$. Then, we can reproduce the
reserves-and-quotas rule, because if there are $\rho _{t}$ or more type $t$
students assigned to a school, then any type $t$ student is unacceptable to
it.

\section*{Appendix D}

In this Appendix, we discuss the running times of our mechanism $\bar{\phi}$%
. Constructing $C_{s}^{\succsim }\left( J\right) $ for all $s\in S$ and all $%
J\subseteq I$ is not finished within a polynomial-time, because there are $%
2^{\left\vert I\right\vert }$ possible subsets of $I$. Thus, to
computationally efficiently execute $\bar{\phi}$, we derive $C_{s}^{\succsim
}\left( J\right) $ as much as needed. Since the SPDA is finished within $%
\left\vert I\right\vert \times \left\vert S\right\vert $ rounds, we need to
derive $C_{s}^{\succsim }\left( J\right) $ for at most $\left\vert
I\right\vert \times \left\vert S\right\vert $ times. Then, we consider the
running time to have $C_{s}^{\succsim }\left( J\right) $ for given $s\in S$
and $J\subseteq I$. First, we have at most $\left\vert I\right\vert $ steps
to construct $C_{s}^{\succsim }\left( J\right) $ by the method above. Let $g$
be a step to construct $C_{s}^{g}\left( J\right) $. If $\left\vert
C_{s}^{g-1}\left( J\right) \right\vert <q_{s}$, we immediately have $%
C_{s}^{g}\left( J\right) $. Thus, we assume $\left\vert C_{s}^{g-1}\left(
J\right) \right\vert =q_{s}$ and choose $r_{g}$ who is most recently applies
to $s$ among the students who have the lowest priority within $%
C_{s}^{g-1}\left( J\right) \cup \left\{ a_{g}\right\} $. Let $J=\left\{
a_{1},\ldots ,a_{\left\vert J\right\vert }\right\} $. To choose $r_{g}$, we
use the algorithm below.

\textbf{Step} $0.$ Let $i_{0}=a_{g}$

\textbf{Step} $\delta =1,\ldots ,g$: If 
\begin{equation*}
i_{\delta -1}\succsim _{s}^{\left( C_{s}^{g-1}\left( J\right) \cup \left\{
a_{g}\right\} \right) \setminus \left\{ i_{\delta -1},a_{\delta }\right\}
}a_{\delta },
\end{equation*}%
then let $i_{\delta }=a_{\delta }$. Otherwise, then let $i_{\delta
-1}=i_{\delta }$.

Then, we let $r_{g}=i_{g},$ who is most recently applies to $s$ among the
students who have the lowest priority within $C_{s}^{g-1}\left( J\right)
\cup \left\{ a_{g}\right\} $. and is rejected in step $g$. Hence deriving $%
C_{s}^{g}\left( J\right) $ is $O\left( \left\vert I\right\vert \right) $ and
constructing $C_{s}^{\succsim }\left( J\right) $ is $O\left( \left\vert
I\right\vert ^{2}\right) $.

Hence, the running time of $\bar{\phi}$ is $O\left( \left\vert I\right\vert
^{4}\left\vert S\right\vert \right) $.

\section*{Appendix E}

Here, we introduce a model with slot-specific priorities \`{a} la Kominers
and S\"{o}nmez (2016). In their model, each school has $q_{s}$ slots denoted
by $\mathrm{s}^{1},$ $\mathrm{s}^{2},\ldots ,\mathrm{s}^{q_{s}}$. Each slot $%
\mathrm{s}^{k}$ has a total order priority denoted by $\Pi ^{\mathrm{s}^{k}}$
for $k=1,2,\ldots ,q_{s}$. For a given $J\subseteq I,$ each school chooses
the subset of $J$ by sequentially filling the slots. That is, the set of
students chosen in the following Steps $1,\ldots ,K$ is the choice of school 
$s$, where $K=\max \left\{ \left\vert J\right\vert ,q_{s}\right\} $.

\begin{description}
\item[\textbf{Step} $k=1,\cdots ,K$] School $s$ chooses $\Pi ^{\mathrm{s}%
^{k}}$-maximal among the students who are in $J$ and have not chosen yet.
\end{description}

We show that any slot-specific priorities may fail to choose the assignment
that maximizes the objective function of school, which is discussed in
Section 5. Suppose that $q_{s}=2$ and thus $s$ has two slots $\mathrm{s}^{1}$
and $\mathrm{s}^{2}$. There are seven students described in the following
table.

\begin{center}
$%
\begin{tabular}{|l|l|l|l|l|l|l|}
\hline
Student $i$ & 1 & 2 & 3 & 4 & 5 & 6 \\ \hline
Type $\tau \left( i\right) $ & $A$ & $A$ & $A$ & $B$ & $B$ & $B$ \\ \hline
Score $\sigma _{i}$ & 1.0 & 1.0 & 0.45 & 1.0 & 1.0 & 0.2 \\ \hline
\end{tabular}%
$
\end{center}

Suppose that the objective function of school $s$ is given by (\ref{c}).
Then, the optimal choice of $s$ is summarized in the following table.

\begin{center}
\begin{tabular}{|l|l|l|l|l|l|l|}
\hline
& Case 1 & Case 2 & Case 3 & Case 4 & Case 5 & Case 6 \\ \hline
Set of applicants $J$ & $\left\{ 1,3,6\right\} $ & $\left\{ 3,4,6\right\} $
& $\left\{ 3,5,6\right\} $ & $\left\{ 3,4,5\right\} $ & $\left\{
2,3,6\right\} $ & $\left\{ 1,2,6\right\} $ \\ \hline
Optimal choice $J^{\ast }$ & $\{1,6\}$ & $\{3,4\}$ & $\{3,5\}$ & $\{4,5\}$ & 
$\{2,6\}$ & $\{1,2\}$ \\ \hline
\end{tabular}
\end{center}

We show that it is impossible to achieve the optimal choice in all cases of
the table above by any slot-specific priorities. To achieve the optimal
choice $J^{\ast }$ of Case 1, either $1\Pi ^{\mathrm{s}^{1}}6$, $1\Pi ^{%
\mathrm{s}^{1}}3$ and $6\Pi ^{\mathrm{s}^{2}}3$ or $6\Pi ^{\mathrm{s}^{1}}1$%
, $6\Pi ^{\mathrm{s}^{1}}3$ and $1\Pi ^{\mathrm{s}^{2}}3$ must be satisfied.

First, we assume $1\Pi ^{\mathrm{s}^{1}}6$, $1\Pi ^{\mathrm{s}^{1}}3$ and $%
6\Pi ^{\mathrm{s}^{2}}3$. To achieve $J^{\ast }$ in Case 2, $3\Pi ^{\mathrm{s%
}^{1}}4$, $3\Pi ^{\mathrm{s}^{1}}6$ and $4\Pi ^{\mathrm{s}^{2}}6$, because $%
6\Pi ^{\mathrm{s}^{2}}3$. To achieve $J^{\ast }$ in Case 3, $3\Pi ^{\mathrm{s%
}^{1}}5$, $3\Pi ^{\mathrm{s}^{1}}6$ and $5\Pi ^{\mathrm{s}^{2}}6$, because $%
6\Pi ^{\mathrm{s}^{2}}3$. Since $3\Pi ^{\mathrm{s}^{1}}4$ and $3\Pi ^{%
\mathrm{s}^{1}}5$, we cannot achieve $J^{\ast }$ in Case 4, but $3$ is
assigned to $\mathrm{s}^{1}$ in that case.

Second, we assume $6\Pi ^{\mathrm{s}^{1}}1$, $6\Pi ^{\mathrm{s}^{1}}3$ and $%
1\Pi ^{\mathrm{s}^{2}}3$. To achieve $J^{\ast }$ in Case 2, $4\Pi ^{\mathrm{s%
}^{1}}3$, $4\Pi ^{\mathrm{s}^{1}}6$ and $3\Pi ^{\mathrm{s}^{2}}6$, because $%
6\Pi ^{\mathrm{s}^{1}}3$. To achieve $J^{\ast }$ in Case 5, $6\Pi ^{\mathrm{s%
}^{1}}2$, $6\Pi ^{\mathrm{s}^{1}}3$ and $2\Pi ^{\mathrm{s}^{2}}3$, because $%
3\Pi ^{\mathrm{s}^{2}}6$. Since $6\Pi ^{\mathrm{s}^{1}}1$ and $6\Pi ^{%
\mathrm{s}^{1}}2$, we cannot achieve $J^{\ast }$ in Case 6, but $6$ is
assigned to $\mathrm{s}^{1}$ in that case.

\end{document}